\documentclass{article}
\usepackage{arxiv}
\usepackage{amssymb}
\usepackage{amsmath}

\newcommand{\shorttitle}{Randomness before Probability: Martin-L\"{o}f randomness of Detailed Data}

\begin{document}

\title{Randomness before Probability, Quantised Gas Laws Directly from
Objective Martin-L\"{o}f Randomness of Detailed Data}
\author{David Sherwell \\
School of Computer Science and Applied Mathematics\\
Universitiy of the Witwatersrand, Johannesburg, South Africa\\
}
\maketitle

\begin{abstract}
We show that objective Martin-L\"{o}f randomness and Kolmogorov complexity 
of instantaneous detailed data lists for $N$ helium gas atoms on $M$ possible 
energies is necessary and sufficient to directly write down its Helmholtz 
free energy and thus all thermostatics of the gas. We show that such 
theory formally \emph{precedes} application of probability and statistics. 
Each datum in a list is distinct if $N,M$ are formally well defined and 
with passive monitoring of thermostatic variables only, each is to be 
intrinsic. In this introductory paper we consider a low density cool gas 
of noninteracting $^{3}\text{He}$ atoms in quantum and classical regimes. 
Objective definitions of detailed disorder and of thermostatic entropy arise 
for gas in spontaneous detailed motion along with new insights into 
irreversible processes and an objective Second Law. Algorithmic probability is 
rigorously associated with Kolmogorov complexity. A condition for equal 
a priori probability naturally arises and Fermi-Dirac quantum statistics 
follows from Martin-L\"{o}f randomness.
\end{abstract}

\section{Introduction}

To outline a new approach to equilibrium quantum fluid theory we briefly
recall properties of objective randomness \cite{LV}. A particular one-way
infinite binary sequence $x\in \{0,1\}^{\infty }$ is random iff it satisfies
all computable tests for randomness \cite[pg220]{LV}. This is the definition
of Martin-L\"{o}f \cite{ML}. Computation is with respect to a universal
Turing machine and Martin-L\"{o}f randomness of a given sequence is not
computably decidable by that machine.\ Kolmogorov prefix complexity $%
K(x_{1:n})$\ is the length of the shortest algorithm that outputs the first $%
n$ digits of prefix-free $x$ because each datum has the same length \cite{LV}. 
$K(x_{1:n})$ is not computably decidable by the universal
Turing machine.\ Martin-L\"{o}f's definition of randomness on infinite
sequences is made serviceable by theorems of Schnorr \cite{Schnorr} and
Levin \cite{Levin}\ based on a proposal of Kolmogorov \cite[pg212]{LV} that
infinite sequence $x$ is Martin-L\"{o}f random (say ML-random) with respect
to uniform measure if and only if \cite[pg215]{LV}%
\begin{equation}
\exists c\forall n[K(x_{1:n})\geq n-c]  \label{1}
\end{equation}%
if and only if
\begin{equation}
\lim_{n\rightarrow \infty }[K(x_{1:n})-n]=\infty .  \label{2}
\end{equation}%
Uniform measure is a natural condition for equal a priori probability of
infinite sequences. Finite binary sequence $x_{1:n}$\ of length $n$\ is 
\emph{incompressible} if \cite[Thm 3.3.1]{LV} 
\begin{equation}
K(x_{1:n})\geq n.  \label{3}
\end{equation}%
Initial subsequences of an infinite ML-random sequence are incompressible.
The set of all incompressible initial subsequences such that $K(x_{1:n})=n$
lies with a mixture of ML-random and nonrandom infinite sequences. Among $%
2^{n}$ possible subsequences of digit length $n$ a fraction $1-2^{-n}$, $%
n\gg 1$\ are incompressible \cite[Example 3.3.3]{LV}. Here, initial
subsequences are \emph{effectively always} incompressible, where it.

Our approach to a theory, strictly of equilibrium fluids, is outlined. We
are given pure thermodynamics $\mathbb{TD}$, say as before publication of
Planck's constant in 1900 \cite{Fermi},\cite{Pipp},\cite{B1}. This is theory
of zero and first order homogeneous (intensive and extensive) functions with
in particular the First and Second Laws. The only measurements permitted are
of variables $T,P,\dots\approx cnst$\ in $\mathbb{TD}$, with passive
monitoring by thermalised thermometers, barometers, ... . Detailed dynamics
is fundamental in quantum statistical mechanics $\mathbb{QSM}$ and the same
holds here.

In detailed physical theory, for a given gas of $N$ noninteracting fermions
in macroscopic volume $V$ closed boundary $\partial V$ in thermodynamic
equilibrium, suppose for a moment the existence of $M\gg N$ discrete
real-valued eigenenergies of one-atom Schr\"{o}dinger problems as might be
spanned by the gas. $N$ free structureless atoms of mass $m$ are distributed
one only per level over these, in accordance with Pauli exclusion. The
numbers $M,N$ are fundamental in $\mathbb{QSM}$ and the same applies here.
In low density gas $M\gg N$ it is long understood that ideal gas and
semiclassical detailed mechanics are consistent with thermodynamics \cite{B1}%
. We keep $^{3}\text{He}$ in mind, like atoms have finite size (ground state
Bohr radius) and mass, are in classical mechanics in $V$ and in the absence
of measurements of any one atom, detailed data here is of particles and is 
\emph{intrinsic}. We show below how this is possible in the face of the
quantum measurement problem (observation disturbs measurements). $M$ distinct 
real-number energy levels are \emph{%
necessarily} defined to at least $k_{M}=\log _{2}M$ digits per datum to
distinguish levels, a \emph{primitive} physical characteristic. For a given
gas conceive an instantaneous list $\mathcal{L}_{M}$ of $M$ discrete
eigenenergies in increasing order and as any one might be computed to
primitive precision. We do not assume that the list itself can be computably 
enumerated. Reversibly encode $\mathcal{L}_{M}$ by concatenation per distinct 
datum to a primitive binary initial subsequence of total digit length 
$l_{M}=M\log M$ ($\log =\log _{2}$). One dynamical property at least of $N$ 
\emph{like atoms} 
must be listable to at least $k_{N}=\log N$\ digits to distinguish atoms and
similarly encode a list $\mathcal{L}_{N}$ to a primitive initial subsequence
of length $l_{N}=N\log N$. Let $\mathcal{L}_{M},\mathcal{L}_{N}$ now denote
finite primitive binary sequences. Numbers $M,N$\ are \emph{well-defined}
if and only if intrinsic primitive lists $\mathcal{L}_{M},\mathcal{L}_{N}$ 
(as finite binary sequences) are physically conceivable.

In $\mathbb{QSM}$ in classical approximation for noninteracting atoms in
equilibrium gas\ 
\begin{equation}  \label{4}
-k_{B}TN\ln Z_{C}(T,V,N)=F(T,V,N),
\end{equation}%
where $Z_{C}=\sum_{i,up}e^{\epsilon _{i}/k_{B}T}$\ is the familiar canonical
partition function and $F$\ is Helmholtz free energy \cite{B1}. Recall $%
e^{\epsilon _{i}/k_{B}T}/Z_{C}$ is probability that an eigenenergy $\epsilon
_{i}$ is sampled from the spectrum of a one-particle Schr\"{o}dinger problem
in a vessel with wall $\partial V$. For spin-1/2 fermions $%
E=\sum\nolimits_{i=1}^{M}\omega _{i}\epsilon _{i}$, $\omega _{i}\in \{0,1\}$
is internal energy, $N=\sum\nolimits_{i=1}^{M}\omega _{i}$, statistical
entropy $S=(E-F)/T$\ is maximised under the two constraints and is
identified as ``disorder''. The fully quantum theory is owing to von Neumann
and we denote statistical entropy $S=S_{vN}$. The $\epsilon _{i}$ are
observables and owing to disturbance of properties by detailed observation,
are not otherwise intrinsic properties.

Fermi-Dirac combinatorics $\binom{M}{N_{s}}$ consistent with Pauli exclusion
and indistinguishability of spin-up fermions underlies $\mathbb{QSM}$ ($%
F=F_{up}+F_{down}$ and $\binom{M}{N_{s}}^{-1}$ is probability that a partial
gas is sampled from the ensemble with cardinality $\sum\nolimits_{s}\binom{M%
}{N_{s}},N_{up}=N_{down}=N/2$). Note the implicit assumption of equal a
priori probability of gas samples. The new approach to $\binom{M}{N_{s}}$ is
not in statistics. With the aid of Stirling's approximation it is known that 
\cite{LV} 
\begin{equation}  \label{5}
\ln \binom{M}{N_{s}}=N_{s}\ln \frac{M-N_{s}}{N_{s}}+M\ln \frac{M}{M-N_{s}}+%
\frac{1}{2}\ln \frac{M}{N_{s}(M-N_{s})}+O(1).
\end{equation}%
Before deciding the proper physical definition of $N_{s}$, in the new theory
we rewrite this for $M>N_{s}$ as%
\begin{equation}  \label{6}
\log \binom{M}{N_{s}}=M\log M-N_{s}\log N_{s}-(M-N_{s})\log (M-N_{s})+\frac{1%
}{2}\log \frac{M}{N_{s}(M-N_{s})}+O(1).
\end{equation}%
In $\mathbb{TD}$\ we note an extensive expression of three terms in
primitive $n\log n$ form of length of detailed intrinsic data lists, and an $%
\sim \log $-term that is neither. We do not consider most probable gases
from an ensemble. Instead we consider a given gas with instantaneous data
lists $\mathcal{L}_{i}$ that change in time and spontaneously sample from
some subspace of cardinality $\sum\nolimits_{s}\log \binom{M}{N_{s}}
$, not defined a priori as the most likely but as a consequence of
persistent underlying ML-randomness. ML-randomness is not computably
decidable and all we can do is make the assumption that the primitive forms
are physically significant finite initial subsequences $\mathcal{L}_{i}$ of
real-valued ML-random data lists $\mathcal{L}_{i}^{\infty }$ on some
subspace of all lists, then test. It is already known from $\mathbb{QSM}$
that empirical ideal gas laws flow from \eqref{4} based on $%
\sum\nolimits_{s}\binom{M}{N_{s}}$ from \eqref{5} and the same must
apply for \eqref{6}. By \eqref{1}\eqref{2} $\mathcal{L}%
_{i}^{\infty }$ are under the assumption almost always ML-random in uniform
measure and we note immediately that \emph{uniform measure admits equal a
priori probability} per instantaneous infinite list that is sampled over
time. If the assumption is justfied by consistency with empirical properties
equal a priori probability can be assumed if $\mathbb{QSM}$ is useful and we
note precedence of the new theory. In \eqref{3} primitive lists $%
\mathcal{L}_{i}$ are then \emph{effectively always} incompressible and in
this sense we interpret the primitive forms \eqref{6} in terms of Kolmogorov
complexities $K(\mathcal{L}_{i})$. For discretised data by some quantum
mechanical process we foresee a link between quantum probability and
ML-randomness which is of fundamental interest. A further contrast is that
in $\mathbb{QSM}$ the statistical entropy $S_{vN}=cnst$\ for exact data (of
detailed underlying dynamics), against the Second Law, and that \emph{coarse
graining of data is necessary} \cite[pg118]{B1}. Here we may consider in a
precise way the exact data.

Consider collisionless gas in a vessel that, however the empty vessel might be filled, is
observed to relax to equilibrium. Because $K(x_{1:n})$\ is an intrinsic
property of any sequence whatsoever, the very general function%
\begin{equation}  \label{7}
D_{t}^{-}(\{K(\mathcal{L}_{it})\}_{i})=K(\mathcal{L}_{Mt})-K(\mathcal{L}%
_{Nt})-K(\mathcal{L}_{M-N,t})
\end{equation}%
is defined at all times for our particular gas (hinting at its relevance to
nonequilibrium gas). For spin-1/2 fermions, for equilibrium thermodynamics
we define the \emph{recursive} (halting-computable \cite{LV}) function of
complexities 
\begin{equation}  \label{8}
D_{s}^{-}(\{K(\mathcal{L}_{i})\}_{i})\simeq M\log M-N_{s}\log
N_{s}-(M-N_{s})\log (M-N_{s})>0.
\end{equation}%
Here $M>N_{s}$\ and $D_{s}^{-}>0$ by \eqref{5}. Note $\simeq $\ implies
that less significant details than $O(MN)$ are ignored for equilibrium and
only extensive terms are significant.

As in $\mathbb{QSM}$ and below let $M=M(T,V)$ \cite{B1}. Define total $%
D^{-}(T,V,N/2)$ by 
\begin{equation}  \label{9}
\sum_{s=up,down}D_{s}^{-}(\{K(\mathcal{L}_{is})\}_{i})\simeq D^{-}(T,V,N).
\end{equation}%
Then for equilibrium spin-1/2 fermionic gas, directly define (Helmholtz)
free energy $F(T,V,N)$ and a function $z_{C}(T,V,N)$ at most significant
precision, by 
\begin{equation}  \label{10}
-k_{B}TN\ln z_{C}(T,V,N/2)=F=-k_{B}TD^{-}(T,V,N)<0
\end{equation}%
In the left equality $z_{C}$ takes the role of, but is \emph{not} formally
equal to, the canonical partition function $Z_{C}$ \eqref{4} of $\mathbb{%
QSM}$, for the right equality $D^{-}(T,V,N)$\ lies with a Massieu function
of $\mathbb{TD}$\ ($\ln z_{C}$ \cite{B1}). By incompressibility, \emph{%
extensive} $D^{-}$\ \emph{is directly physically maximal} (no need for maximising under
constraints as above) and $F$\ is minimal as expected for equilibrium in a
potential. By \eqref{8} the three \emph{numbers} $K(\mathcal{L}%
_{n})= nlog{n}$ are \emph{new physical variables} in equilibrium gas
theory. Approach from the right of \eqref{10} is sufficient, with the
definitions \eqref{6} \eqref{8}, to entirely capture thermostatics of
a fermionic gas. Granted data lists, theory on the right of \eqref{10}
is closed and is then physically independent of theory on the left. We refer
to theory strictly on the right as \emph{quantised complexity mechanics} $%
\mathbb{QKM}$. At equilibrium we refer to \emph{detailed net disorder} $D^{-}
$, ``disorder'' expressing the only direct reference to ML-randomness. We
develop this below and return to theory of the left equality only in a
finalising section \S 6 where we touch on algorithmic probability (from $1/%
\binom{M}{N}\sim 2^{-D^{-}}$). It too is closed, lies with data lists, is
complementary to $\mathbb{QKM}$ on the right but is qualitatively distinct,
and, is qualitatively distinct from $\mathbb{QSM}$.

In $\mathbb{QKM}$\ one at least of the three lists of \eqref{7} is \emph{%
physically independent} of the others (because any two interrelated terms
can be used to shorten the algorithm for the third, contradicting %
\eqref{8}). So not all three can be lists of eigenenergies, as expected. The
first and third are instantaneously of all energies sampled and all
unsampled energies, the second is of particles but is not of their energies.
Not only does $\mathbb{QSM}$\ in its sole use of the spectrum $\epsilon (%
\mathbb{SM},m,\partial V)$ purposely ignore physical properties of data by
directly turning to probability \cite{B1}, but the independence of lists
shows that it purposely \emph{ignores objective physical mechanism} at a
fundamental level. Although thermostatics follows from \eqref{10} \emph{%
relaxation} to equilibrium and \emph{persistence} of equilibrium is observed
data, and, is deeply related to the Second Law (\S 5). For Fermi-Dirac
combinatorics these three phenomena are simultaneously effective. We account
for relaxation and persistence with the help of these hidden mechanisms.%
The emergence of additional physics for
equilibrium fluid highlights that \emph{randomness lies before probability}.
Physics through the new integer variables $K(\mathcal{L}_{i})$, not
statistics, takes precedence.

It is striking that $K(\mathcal{L}_{M})\simeq M\log M$ is accounted by
effectively always incompressibility of energy data for almost always
ML-random energy levels. Consider the introductory textbook demonstration of 
$\mathbb{QSM}$ for noninteracting gas in a smooth box of volume $V$ \cite{B1}%
. It is implicit that the box has macroscopic size far exceeding quantum
atomic scales. For the box there is a short $O(\log M)$ formula fixed by $%
\partial V_{box}$ to list these and they are \emph{not} incompressible. Its
real-valued quantum spectra are not ML-random and in turn it follows that
equal a priori probability fails. $\mathbb{QSM}$ is not applicable. In $%
\mathbb{QSM}$\ the problem $\epsilon (\mathbb{SM},m,\partial V_{box})$ for
the spectrum of the smooth box is presented as adequate to derive state
equations where the environment of the gas (the smooth box) is trivially
represented by a simple closed wall of unspecified temperature. All this is
unsatisfactory. The density operators of $\mathbb{QSM}$\ are partly
objective owing to quantum uncertainty, and we respect their relevance for
equilibria, but as $\binom{M}{N}$ is used they hide detailed physics. In
particular, evolution of gas to equilibrium is passed to the Liouville-von
Neumann equation and Ehrenfest equation for evolution of operators in
quantum mechanics, and to Boltzmann's equation in classical mechanics \cite[%
Sections 2.3 and 3.3]{B1}. We formally relate equilibrium $\mathbb{QKM},%
\mathbb{\mathbb{QSM}}$\ in \S 6.

We study a given gas. We differ from all derivations of equations of state
for fluids based on classical or quantum statistics over an ensemble of
fluids. To our knowledge, this is all contemporary gas and liquid studies.
The widely held present view that thermodynamic entropy is a specialised
form of Shannon statistical entropy in information theory and is a measure
of ignorance lies with ensembles \cite{B1}. Interpretation of ``disorder" as
introduced by Clausius (1862) is his irreversible entropy with inevitable
decay of order \cite{B1}, but \emph{objective} disorder, a certain precise
knowledge, by any definition is not yet established. Nor does equal a priori
probability have an objective footing. Application of Kolmogorov complexity
has been considered as a natural measure of disorder but in the context of
ensembles and in the limited case of a single data list \cite{LV},\cite{Zur},%
\cite{V}. Brudno's theorem relates Kolmogorov complexity rate $%
\lim_{n\rightarrow \infty }\frac{1}{n}K(x_{1:n})$ to Kolmogorov-Sinai
KS-entropy measure in ergodic theory of classical dynamical systems and has
been extended to rate of von Neumann entropy production for quantum
ensembles, both notably hint at homogeneity at equilibrium but a single $%
K(x_{1:n})$ by itself is, we have shown, not sufficient for thermostatics 
\cite{Brud},\cite{Benatti}.

\section{Thermal length, collisionality}

In natural variables $T,V,N$ with fundamental constants $m,k_{B},h$\ the
only nondimensional length that can be defined starting from $T$ is $\lambda
_{th}$\ defined by 
\begin{equation}  \label{11}
\lambda _{th}^{2}=\frac{h^{2}}{2\pi mk_{B}T}.
\end{equation}%
where a factor $2\pi $\ aligns with $\mathbb{QSM}$ (and has further
adjustment for objective physics in \S 6). Let $N=N_{up}+N_{down}$. The
simplest such \emph{intensive} parameter for given $N,V$ is 
\begin{equation}  \label{12}
A=\frac{V}{N\lambda _{th}^{3}}=\frac{M}{N}\text{, }M=\frac{V}{\lambda
_{th}^{3}}=\gamma VT^{3/2}.
\end{equation}%
In $\mathbb{QSM}$\ the standard interpretation of $\lambda _{th}$ is in de
Broglie's sense of wavelength of the typical particle, $p_{th}=h/\lambda
_{th}=\sqrt{2m\epsilon _{th}}$\ is typical momentum and $M=V(2m\epsilon
_{th}/h^{2})^{3/2}$ is interpreted as the effective number of discrete
energy levels spanned \cite{B1}. Note 
\begin{equation}  \label{13}
A=\left(\frac{\ell _{N}}{\lambda _{th}}\right)^{3}\text{, }\ell _{N}=\left(\frac{V}{N}\right)^{1/3}%
\text{, }b=V^{1/3}
\end{equation}%
where $V/N$\ is inverse density so that in present theory $A$ is no more
than the intensive ratio of macroscopic scales to microscopic scales $%
\lambda _{th}$, where the latter are to characterise detailed physics. The 
interpretation $A^{-1}=N/M$ as a measure of the number of atoms per energy
level is not found to be useful here (see \S 6).

There are six length scales of interest for a collisionless $^{3}\text{He}$\
gas. We have spherical ground state orbital at Bohr radius $a_{B}=0.53\times
10^{-10}m$ from quantum theory and experiment \cite{B1}. Concerning
interactions, the empirical Lennard-Jones potential for helium 
\begin{equation}\label{14}
\psi _{LJ} =4\varepsilon \left(\left(\frac{a_{LJ}}{r}\right)^{12}-\left(\frac{a_{LJ}}{r}\right)^{6}\right),
\end{equation}
\begin{equation}\label{15}
\varepsilon \approx 11\text{, }a_{LJ}\approx 2.6\mathring{A}>2a_{B}=1.1\mathring{A}
\end{equation}
models closeness of approach of two interacting molecules, a very strong
exclusion from below about $a_{LJ}=2.6\mathring{A}\approx 2.6(2a_{B})$ and
an ineffective quantum trap at greater separations \cite{B1}. In the absence
of bonding of $^{3}\text{He}$ Schr\"{o}dinger mechanics 
is certainly effective on scales up to $\sim$ $a_{LJ}$.

Detailed discussion of conditions for noninteracting ideal gas, and for
classical detailed mechanics is given in \cite[pg311]{B1} and will apply
here for gas in $V$. In homogeneous fluid classical collision free path $%
l_{mfp}=V/\sqrt{2}\pi Na^{2}$\ introduces a sixth length scale where $a^{2}$%
\ is cross section and $a$ is impact parameter for a hard surface \cite{B1}.
In quantum mechanics detailed cross sections remain useful \cite{Schiff}.
Gases are \emph{collisionless/collisional} for mean free path $%
l_{mfp}\gtrless b$. In a follow-up paper we will test $\mathbb{QKM}$ against
the gravity-free experiments of \cite{Lipa1},\cite{Lipa2} for specific heat
at the $\lambda $-point in collisional liquid $^{4}\text{He}$. There $b=3.5cm
$ for a spherical copper calorimeter and we conveniently introduce method
for collisionless classical $^{3}\text{He}$ gas in this device.\ Assuming
this value for $b$ and for $a_{LJ}>2a_{B}$ using $a_{LJ}=2.6\times 10^{-10}m$
as an estimate for cross section, helium gas is \emph{collisionless} if $%
l_{mfp}>b$ or $N=V/\sqrt{2}\pi la_{LJ}^{2}<1.7\times 10^{16}$, $2.8\times
10^{-8}mole$ in \emph{gas phase}. Length scales in such low-density gas
satisfy $\ell _{N}=\allowbreak 2.2\times 10^{-7}m\gg a_{LJ}>a_{B}$ so that
neither quantum interaction length is effective. For $^{3}\text{He}$ $%
T_{crit}=3.3K$ and we consider \emph{cool} gas, say $3.3K<T<10K$ which
admits a testable small quantum correction from fully quantised state
equations from \eqref{10}. We find $5.7\mathring{A}>\lambda _{th}>3.1%
\mathring{A}\gtrsim a_{LJ}$. For $T=10K$, $N=\allowbreak 1.7\times 10^{16}$, 
$P=Nk_{B}T/V=\allowbreak 1.3\times 10^{-2}Nm=\allowbreak 1.3\times 10^{-7}bar
$. Thermal velocity is $v_{th}=290ms^{-1}$, $k_{B}T=0.88m\,eV$. Finally $%
A=(\ell _{N}/\lambda _{th})^{3}=3.6\times 10^{8}\gg 1$, well above observed
quantum phenomena where $A\sim 1$ at $T\sim 2.5mK$ \cite{Osch}. Here $%
M=6.1\times 10^{24}$. By $A\gg 1$ we have a \emph{cool collisionless ideal} $%
^{3}\text{He}$\ gas. To summarise example gas properties, in the device of 
\cite{Lipa2} $T,M,N,A$\ are respectively 
\begin{equation}\label{16}
10K,6.1\times 10^{24},1.7\times 10^{16},3.6\times 10^{8}.
\end{equation}

As $T\rightarrow T_{crit}=3.3K$\ $\lambda _{th}$\ increases which is a \emph{%
decoupling} from detailed quantum scales $a_{LJ}$, to be re-interpreted in
equilibrium thermodynamics (\S 6), not in quantum mechanics. In summary, for
collisionless gas 
\begin{equation}\label{17}
l_{mfp}\geq b>\ell _{N}>\lambda _{th}\gtrsim a_{LJ}>2a_{B}\text{, }%
T_{crit}<T\lesssim 10K.
\end{equation}%
Experiments and quantum mechanics of $^{3}\text{He}$ have been of interest
at very high pressures $\sim 30bar$, $T\sim 2.5mK$, in particular in regard
to its bosonic nature where $A\lesssim 1$ \cite{Osch},\cite{Leg}. Very low
density gasses will serve particular physical effects of ML-randomness and
Kolmogorov complexity alone.

We finalise total detailed net disorder for $^{3}\text{He}$ equilibrium of
mixtures of spins. For thermodynamic equilibrium of two spin species, for
all $A=M/N>1/2$, $D^{-}=D_{up}^{-}+D_{down}^{-}$, 
\begin{align}
D^{-} &=2\ln 2(M\log M-N/2\log N/2-(M-N/2)\log (M-N/2)) \label{18} \\
&=N[\ln (2A-1)-2A\ln (1-(2A)^{-1})]\text{, }A=M/N>1/2. \label{19}
\end{align}%
For $A\gg 1$, in the classical limit we find 
\begin{equation} \label{20}
D_{cl}^{-}=N(\ln M-\ln N/2+1+(2A)^{-1}+O(A^{-2}))
\end{equation}%
which agrees with $\mathbb{QSM}$ to $O(A^{-1})$\ and is a first test \cite[%
pg56]{B2}. It is striking that for $A=M/N\ge 1$ one atom per distinct energy level 
occurs satisfying Pauli exclusion. For $1>A\ge {1/2}$ pairs of up/down spins 
per level are mixed with singletons. For $A<{1/2}$ $D^{-}$ is not defined and 
fermions can no longer be supported at all. This structure agrees with 
$\mathbb{QSM}$ where it is defined \cite{B2}. 

A population of $N$ spin-zero bosons where the quantum correction
is negative does not agree with $D^{-}$ but can be shown to agree with $%
\mathbb{QSM}$ for Bose-Einstein combinatorics for $M,N>0$ given by 
\begin{equation} \label{21}
\log \binom{M+N}{N}=(M+N)\log (M+N)-M\log M-N\log N+\frac{1}{2}\log \frac{M+N%
}{MN}+O(1).
\end{equation}%
Analogously, it has a significant extensive part $D^{+}$ and a nonextensive
correction. In the limit $M/N\gg 1$\ the two quantum statistics agree to $%
\sim A^{0}$ and $D^{-}$ will suffice in order to demonstrate method in $%
\mathbb{QKM}$. $^{4}\text{He}$ is spin zero and there $N$\ atoms are
indistinguishable. Of course above fermion pairs are $N/2$ bosons and (21) 
will apply for $0<{2A}<1$, that is for $M/(N/2)<1$ and arbitrary stacking 
of pairs per energy level. 

These two combinatorics \eqref{6}\eqref{21} have an extensive
significant part which is \emph{perfect}, by which we mean that their
logarithm is expanded in primitive $n\ln n$ terms of detailed data to $%
O(\log MN)$. By consistent success against experiment, single-particle
quantum mechanics with $\mathbb{QSM}$\ has discovered perfect 
$\log $-FD or $\log $-BE\ Massieu functions that are related to intrinsic
quantum probability and statistics. Entropy as developed in $\mathbb{QKM}$
from the closed problem on the right of \eqref{10}) will no longer be
statistical.\ 

\section{Existance and characterisation of intrinsic detailed data lists for
equilibrium gas}

We are given a particular gas with some set of thermodynamic properties eg $%
T,P,N,...$ and empirical gas laws denoted $\mathfrak{E}_{0}$ in the case of
ideal gas. Equilibrium itself is a datum \emph{monitored} by steady readings
of thermalised macroscopic devices built into the wall which, in experiment,
is a counter-balancing heat- and mass-bath (thermostat and manostat [15]). 
Thermodynamic phase (gas, liquid, saturated, ...) of the given gas is
known data.\ These fix the macroscopic state. Atomic properties such as
translation of atoms with mass, spin, electron shell structure fix detailed
given properties of atomic species. The fundamental constants are assumed to
be given real numbers. Interchange of units has simple rules. Altogether
these simple properties define a trivial list of a few lines.

$K(\mathcal{L}_{n})$ is independent of units and of scale factors of
detailed data (note different underlying units of lists in \eqref{7}),
is independent of coordinate transformations by any given computable
formula, similarly for data over experimental devices, holds for data in
classical (including general relativity) or quantum mechanics and of course
may be proposed for any sequence in any application whatsoever. We
refer to $\mathcal{L}_{n}$ in general as a \emph{pattern} and to the new
variable $K(\mathcal{L}_{n})$ as \emph{quantity of that pattern}. The
variable $0 \lesssim K(\mathcal{L}_{n})\lesssim l(\mathcal{L}_{n})$ is an
element in an equivalence class $[K(y)]$ of instantaneous finite patterns $%
y\in \mathbb{N}$ of equal Kolmogorov complexity. In equilibrium gas and as 
detailed data changes, for infinite/finite lists, almost all/effectively all are 
in the same equivalence class. Here it substitutes for sample space in a 
statistical approach.

The Invariance Theorem \cite{LV} states that Kolmogorov complexity agrees 
across Turing machines up to an additive constant $c$. It is objective but not absolute 
for the single reason that it is defined relative to a particular universal Turing machine. 
It is known that for one such $K(x)$ $c=1066$, we standardise on this machine and assign 
$c$ to the trivial data list so that $K(\mathcal{L} _{i})$ is then absolute in this
gas theory. By this quality, gas properties depend only on detailed data 
$\mathcal{L}_{i}$ itself and nothing else. The new variables are dynamically 
fundamentally coupled to intrinsic detailed data such as position and momentum 
of atoms in classical mechanics. From \eqref{10} the potential to do work at constant 
temperature spontaneously adapts to detailed data alone, as follows.

Consider mathematical physics as a description of data. Pure thermodynamics 
$\mathbb{TD}$ in the form of \eqref{10} invokes $D_{t}^{-}$ \eqref{7} and $D_{s}^{-}$ \eqref{8}, 
both in terms of absolute properties of data lists. Refer to $K(\mathcal{L}_{i})$ 
as \emph{description length} of data. Then \eqref{8} requires no more than 
description length by \eqref{3}, strictly with equality and steady state. It follows that a 
$\mathcal{L}_{i}$ may be an initial sublist of nonrandom 
infinite lists. This may arise for lists if quantum energy levels of a Schr\"{o}dinger 
problem with Dirichlet boundary conditions as follows. 
It is known that every eigenvalue of the Schr\"{o}dinger wave equation in $\partial V$ 
is a computable real number, certainly for example to $MlogM$ digits \cite{PER}.
For any given (not necessarily minimal) algorithm of length $MlogM$ for an eigenvalue,
the wall is defined by a subroutine in the algorithm and the algorithm for
any other level shares that subroutine. In consequence $K(\mathcal{L}%
_{M})<M\ln M$ and they are not ML-random. Refer to \emph{algorithmic correlation} 
of atoms by the wall. Consistently the smooth box has a 
common short $O(\log M)$ formula for each of $M$ levels and levels are 
not ML-random. This is not unexpected for the 
deterministic Schr\"{o}dinger equation. Of course it is still the case that probability of 
observation of an atom at a particular location in $V$ will be determined under 
Born's interpretation of quantum probability \cite{Schiff}. However the measure of lists 
of nonrandom such descriptive lengths is $2^{-l(\mathcal{L}_{M})}\approx 0$. 
These lists are initial sublists of real valued eigenvalues. We reject this 
Schr\"{o}dinger problem for the energy spectrum for gas in greater $\partial V$. A 
converse is that such a physically relevant Schr\"odinger problem is a theory of 
structure in some sense (see \S 7). 

Equally fundamentally, the eigenenergies are quantum observables, are net
of interaction with a measuring device, are not intrinsic in our
experimental thermodynamics and do not define intrinsic lists. At least in 
the present application, Schr\"odinger mechanics in $\partial V$ presents 
structure underlying observations, intrinsic values are fundamentally 
uncertain as quantified by Heisenberg. 

The property of almost always ML-randomness is strong in that any single
intrinsic datum in an ML-random list must be independent of the others. In 
detailed physics atoms see the detailed wall and we consider possibly 
independent interactions on the scale of the copper lattice, say for 
the experiments of \cite{Lipa1}\cite{Lipa2}.  
These might be variously owing to local potential
wells as in adsorption, simple mechanical trapping in the lattice, and must
account for detailed energy exchange with a real, thermalising wall as heat
bath. Complete statistical decorrelation of equilibrium atom energies by
such mechanisms has been known since the adsorption experiments of Knudsen
(from 1909) and quantised theory of Langmuir (1921) and in general
statistical theory including a thermalised wall by for example Fowler (1927) 
\cite[1936, pg698]{Fowler}. Certainly Dirichlet boundary conditions do not 
apply with the further objection to the above 
Schr\"{o}dinger problem that discreteness of levels is no longer guaranteed. 
None of the lists $\mathcal{L}_{M},\mathcal{L}_{N_{s}},\mathcal{L}_{M-N_{s}}$\ is so far
well-defined. 

Recall Langmuir adsorption theory, a balance in thermostatics of atoms in an
ideal gas ($A\gg 1$ is satisfied) and atoms trapped in vacant lattice sites
in the wall with potential $-u$ localised on scales of lattice spacing. It
is an explicit condition that one atom at a time and independently of its
spin is trapped which is consistent with the collisionless gas. In the
canonical ensemble of $\mathbb{QSM}$, for $\mathcal{M}$\ sites on the copper
wall and $\mathcal{N}$\ trapped atoms, one only per site, we find the
Massieu function ($\varphi _{\text{Lang}}=\ln Z_{C,Lang}$) \cite[pg174]{B1}\
\ 
\begin{align}
\varphi _{\text{Lang}} &=\ln \binom{\mathcal{M}}{\mathcal{N}}+\mathcal{N}%
\frac{u}{ k_{B}T}\simeq \lbrack \mathcal{M}\ln \mathcal{M}-\mathcal{N}\ln 
\mathcal{N}-( \mathcal{M}-\mathcal{N})\ln (\mathcal{M}-\mathcal{N})]+%
\mathcal{N}\frac{u}{k_{B}T} \label{22} \\
&=D_{\text{Lang}}^{-}/\ln 2+\mathcal{N}\frac{u}{k_{B}T}\text{, }\mathcal{M}> 
\mathcal{N}. \label{23}
\end{align}%
Here $\varphi _{\text{Lang}}$ is defined only where $u\neq 0$, spatial
scales in lattice $\sim 3\mathring{A}$. In $D_{\text{Lang}}^{-}$\ we
recognise the perfect $\log $-FD combinatorics \eqref{6} rewritten for
Kolmogorov complexity of three incompressible lists denoted $\mathcal{L}%
_{i}^{\prime }$. So we have $\mathcal{M}$ independent sites as required and
ML-random sampling by $\mathcal{N}$\ atoms. The fourth term is extensive, is
a single number and characterises the strength of the trapping potential. $%
\mathbb{TD}$ separates combinatorics for a gas in an external potential by
passing the fourth term to a thermodynamic potential, here to the free
energy as in $F-u$ in \eqref{10}. The fraction of occupied sites
(Langmuir adsorption isotherm) in the wall is%
\begin{align}
\mathcal{N}/\mathcal{M} &=P/(P+P_{0}e^{-u/k_{B}T})=1/(1+Ae^{-u/k_{B}T}), \label{24} \\
P_{0} &=(k_{B}T)^{5/2}(2\pi m/h^{2})^{3/2}, \label{25}
\end{align}%
where in ideal gasses $P/P_{0}=A^{-1}$. In copper we find measured $u\sim
3meV$ for $(3/2)k_{B}T\sim 1meV$\ $A=(3.2\times 10^{8})$ just above
liquidisation of $^{3}\text{He}$ and $\mathcal{N}/\mathcal{M}=\allowbreak
5.6\times 10^{-8}\ll 1$ in collisionless gas \cite{Zarem}. This is a
theory of the gas in the wall. The thermalised copper lattice is
characterised as usual in solid state quantum mechanics and as a heat and
particle bath is regarded as unaffected by light helium atoms, but by its
thermal vibrations is responsible for detailed energy exchange per helium
atom consistently with temperature $T$. We
do not consider quantum mechanics of the copper with helium impurities itself, 
nor the diffusion of helium in copper as heat/particle bath. A \emph{balanced and 
thermalising return flux} is assumed \cite{B1},\cite{Fowler}. There is no detailed
particle balance (as one atom adheres another is not directly emitted). Langmuir's 
Massieu function separates off the equilibrium thermodynamics of trapped helium. 
We recognize uncorrelated one-by-one independent emission from the lattice. We 
will consider also uncorrelated directions of propagation into $V$. The intrinsic property of
lists $\mathcal{L}_{M},\mathcal{L}_{M-N_{s}}$\ remains unclear, $\mathcal{L}_{N_{s}}$
\ is so far undefined. The adsorbed fraction appears small. It is useful to first 
characterise lists in $V$ and then extend discussion of decorrelation in the wall.

Consider atoms in $V$. It is known that for $2A\gg 1$\ the noninteracting
gas in $V$\ is well-described in classical statistical mechanics. A
semi-classical approach is well-proven and standard, with a warning only for 
$^{3}\text{He}$ at $T \sim 10^{-3}K$ \cite{B1}. In a transit
of $V=b^{3}$, $t_{b}=b/p_{th}$, $p_{th}b\gg h$ \ $\epsilon _{th}t_{b}\gg h$\
and the conventional criteria for classical mechanics of point particles in $%
V$ are satisfied, additionally, particles are noninteracting so the
condition $2A=2(\ell _{N}/\lambda _{th})^{3}=7.2\times 10^{8}\gg 1$ is
satisfied even at low $T>T_{crit}$, also it will be found that chemical
potential satisfies $-\mu /k_{B}T=\ln 2A\approx \allowbreak 20\gg 1$ in
regard to particle exchange in classical systems \cite{B1}. Monitoring of
thermostatic properties in $V$\ does not significantly disturb those
properties and we identify the list of data of emitted atoms $\mathcal{L}%
_{M} $ \eqref{7}\ as \emph{intrinsic} in free gas.\ Atoms have finite
size (Bohr diameter $a_{B}=1.2\mathring{A}$), mass localised to this
diameter and we interpret particle, not wave, state. We strengthen this
property. By the fundamental property of Pauli exclusion and by the
discovery of the physical fact of Fermi-Dirac combinatorics \eqref{6}
and statistics (see \S 6) along with the hidden properties of independent
lists $\mathcal{L}_{i}$ in $V$, classical data per atom is independent. The
physical significance of primitive lists admits that in spite of
incomputability of complexity of initial subsequences these can be written
down. These properties are necessary (by \eqref{6}) and sufficient
(there do not need to be longer algorithms) in order to write down
thermostatic equations of state (next section). For $A\gg 1$\ suppose that a
quantum wave packet on scales of $\lambda _{th}$ applies for an atom as it
exits in a site and that classical mechanics applies in the sense of Ehrenfest
theorem, that is so long as there is insignificant spreading of the packet 
\cite{Schiff}. For this very narrow initial packet and at thermal velocities
we find that the spread of a Gaussian packet is $\Delta r=ht_{b}/2m\lambda
_{th}=b/2$ in a transit time so that wave functions typically overlap, they
are not independent and thus contradict the hidden processes of Fermi-Dirac
combinatorics \eqref{6} \cite{Schiff}. It is suggested that classical
mechanics is \emph{not an approximation of quantum mechanics} in $V$.
Expanding wave fronts begin to see common sections of the wall and are
algorithmically correlated, consistently with the above rejection of a Schr%
\"{o}dinger problem in greater $V$. Quantum mechanics does not apply there
at all and detailed data in $V$ is intrinsic. 
In particular $\mathcal{L}_{M}$\ in $V$ is well-defined.
Concerning $\mathcal{L}_{M-N_{\text s}}$\ \eqref{7} the direct
interpretation is of a list of $M-N_{s}$\ distinct unoccupied energy states
as they might be delivered from the wall and is intrinsic, if $\mathcal{L}%
_{N_{s}}$ is found to be independent and intrinsic.

Concerning $\mathcal{L}_{N_{s}}$\ in $V$, magnitude of momentum of atoms is not
independent of energy in classical mechanics and is of no interest in
relation to particle trajectories, but $N$ detailed directions of
propagation $\{\theta _{js}\}$ are intrinsic in $V$ and can be instantaneously
listed as can say $N_{s}$ nearest unassigned neighbour locations $\{d_{js}\}$ of
atoms (each a property of direction cosines; for vector variables, if a list
of a component is not incompressible then neither are these lists of
scalars). $\mathcal{L}_{M-N_{s}}$ in $V$\ is then well-defined from 
$\{\theta _{js}\}$.
Nearest neighbour displacements in $V$ exclude the wall (these are free
atoms), define $\mathcal{L}_{N_{s}}$\, by necessary ML-randomness and incompressibility $\ell
_{N_{s}}=(V/N_{s})^{1/3}$ is well-defined and thus \emph{ensures homogeneity} in $V$.
Incompressible instantaneous lists of $N_{s}$ intrinsic orientantions $\mathcal{L%
}_{\theta s }$ in $V$ are defined which \emph{ensures isotropy }in space and
momentum. $D^{-}=D_{up}^{-}+D_{down}^{-}$ \eqref{19} in $V$ is finally
well-defined along with free energy $F$ \eqref{10}.

Interpret $D_{\text{Lang}}^{-}$\ in $\mathbb{QKM}$. Sites are mechanical 
defects in metal which are not
affected by local helium atom interactions, are intrinsic (a detailed
observation of a defect by scattering a few helium atoms will not affect its
future ability to adhere atoms) and so in regard to their positions in the
wall $K(\mathcal{L}_{\mathcal{M}}^{\prime }),K(\mathcal{L}_{\mathcal{M-N}%
}^{\prime })$ are well-defined if $\mathcal{L}_{\mathcal{N}}^{\prime }$ is
well-defined. As for gas in $V$ we assume ML-randomness of nearest neighbour 
site locations in the wall. Occupied sites $\mathcal{L}_{%
\mathcal{N}}^{\prime }$ are not independent of these two and we use 
complexity of exit orientations $\mathcal{L}_{\theta }^{\prime }$ instead 
(both spins). All atoms in $V$ have interacted with the wall and have 
accumulated to define $\mathcal{L}_{\theta}$. By emission of one atom only 
per site, by mapping of trajectories back to the wall and the necessary 
constancy of complexities in equilibrium we assign intrinsic $\mathcal{L%
}_{\theta }$ to define a list of invariant complexity of exit 
orientations at the wall, notably without time stamp. Time of arrival at 
the wall is ML-random so that no inhomogeneous neighbourhood occurs, 
conversely time of exit from a site must be ML-random for the same reason. 
Both are unacceptable correlations. Detailed instantaneous lists of 
orientations and of exit times from 
occupied sites are conceivable but are not quantitatively pertinent to 
equilibrium. Only their existence is required. Langmuir's assumption of 
one atom only per adhesion site substitutes for quantum mechanical 
Pauli exclusion, justifying perfect $\log $-FD Massieu functions 
in \eqref{23}. The nature of quantum mechancs in a site arises. We address 
algorithmic probability in $\S 6$ and return to mechanics in a site in $\S 7$.

Magnitude of momentum and energy of emitted atoms in the population at the
wall are \emph{not} mentioned in $D_{\text{Lang}}^{-}$. For insight, for
classical mechanics over $V$ recall the classical dynamical system of
elastic scattering of hard spheres at a closed rigid wall of piecewise
positive convex-inward curvature. This is a Sinai billiard which rigorously
is ergodic, mixing and has KS-entropy, together guarantees of homogeneity,
isotropy with exponentially fast separation of nearby atoms (the latter
implies the former two) \cite{SpringBun}. Overall refer to complete
ergodicity. The set of all possible infinite piece-wise paths between
bounces on the wall is \emph{independent of speed} along the paths,
consistently with our preference for orientations $\{\theta _{j}\}$ as
physically significant in $V$ and, is independent of particular shape of
otherwise convex inward walls. Thermal fluctuations of the copper wall
determine instantaneous fine-scale convex inward wall shape, the Sinai
billiard is instantaneously redefined and for short contact times complete
ergodicity persists. \emph{Inelastic} interaction (ie with energy exchange)
with a thermalising wall (heat bath) may be introduced but does not affect
complete ergodicity. The chosen lists are consistent with complete
ergodicity in $V$. The smooth box is not classically mixing \cite{SpringBun}%
. Combinatorics as a foundation for probability always proceeds 
with an assumption of a well-mixed
sample space (eg white and black beans blind sampled from a box) and we note
that combinatorics should not be applied to gases in the smooth box (zero
wall curvature, no mixing, no a priori equal probability). In objective
physics, mixing must be spontaneous and indeed is explicitly demanded by 
Fermi-Dirac combinatorics.

Again refer to the Sinai billiard. The wall of the smooth box might be
deformed to six inward-curving smooth walls intersecting on lines of
singular curvature. The dynamics is completely ergodic, classical
trajectories incident on these lines are \emph{permanently trapped} but have
zero Lebesgue measure \cite{SpringBun}. For atomic-scale curvature the
copper lattice opens and a finite measure of trajectories is mechanically
trapped, independently of adsorption. Again decorrelation is absolute. For
the vessel of \cite{Lipa1}, for lattice spacing $\sim 3\mathring{A}$ in
copper with atom size $1.2\mathring{A}$ a classical helium atom can
penetrate the lattice. As for Langmuir's $-u$ particle dynamics on
microscopic scales $\sim 3\mathring{A}$\ is quantised in a local well in
copper. We model such trapping by substituting $-u/k_{B}T$ with $%
-(n_{latt}-1)$\ where $n_{latt}$\ is the number of contacts of a single atom 
within a copper lattice spacing, $n_{latt}=1$\ represents a directly 
scattered atom. There are again the two populations. Energy transfer by 
quantum thermal fluctuations of copper wall occurs. We assume 
trapping on scales $\lesssim 3\mathring{A}$, that is for 
trapped atoms and for atoms that do a second bounce with a neigbouring copper 
atom then escape. We distinguish a single bounce when $n_{latt}=1$ for single
scatter to a remote wall atom. As for adsorption we assume complete 
ergodicity in $V$. Adapting the Langmuir equilibrium model, for $n_{latt}-1>0$ equivalent to 
$\left\vert u\right\vert /k_{B}T>1$, $\mathcal{N}/\mathcal{M}=1/(1+Ae^{-(n_{latt}-1)})%
\ll 1$ again (for $A$, for ideal gas equilibrium $V/N={V}^{\prime }/{N}^{\prime }$). 
Regarding $n_{latt}$ as a measure of depth of penetration in an effective local well, 
decorrelation can in principle be significant. These Langmuir-type processes 
do not conserve energy at detailed level.

Suppose $u\approx 0$ and/or $n_{latt}=1$\ but a wall of ML-randomly packed
convex-inward copper atoms. Suppose a scattering rule, for example that on
first bounce angle-in equals angle-out (specular reflection with conservation 
of energy if contact time with copper is very short). Then on next
specular reflection on a remote ML-randomly located atom across the vessel,
the first incoming orientation is decoupled from the second outgoing 
orientation, ML-randomly arranged wall atoms imply a complicated
mathematical description and algorithms to locate the next scattering
atom are incompressible. Then we may again assume complete ergodic
properties in $V$ but with understanding that decorrelations take place on
timescales $\gtrsim 2t_{b}\sim 2\times 10^{-4}s$. This will take place even
if energy is conserved. This sets a relaxation time for an initial
nonequilibrium filling gas. Overall we refer to \emph{Langmuir-type}
processes of atoms in and at sites. Altogether they drive observed \emph{%
relaxation} by maintaining independence of atoms. Let $\mathcal{M}_{tot}$\
detailed sites cover the wall. Equilibrium number density in $V$ is
maintained by the real wall as a manostat \cite{B1}. Equal chemical
potentials of gas in $V$ and gas at the wall $\cup \partial V^{\prime }$
ensure $N=cnst$. We find a mix of helium thermodynamics and detailed dynamics 
on quantum scales of a copper lattice. 

Thermal length $\lambda _{th}$ characterises detailed effective scale in the 
lattice. As a heat bath atoms are thermalised and by \eqref{11} we associate 
magnitude of momentum $p_{th}=h/\lambda_{th}$ with the thermalised energy 
$p_{th}^2/2m$. We do not interpret de Broglie wavelength of an atom. 
It follows that a helium atom is localised within precision $\lambda _{th}$ 
in a classical phase volume $h^3$. By the former a wave function as a 
probability amplitude of location in V is redundant while atom energy is 
determined by thermalisation. By classical phase volume $h^3$ Heisenberg 
uncertainty relations hold at their margin so that conjugate variables of 
vector momentum and time in a site are formally unpredictable. We do not 
observe so make no call on a wave function, but we do assume that quantum 
indeterminism is intrinsic, is owing to and is characterised by 
Heisenberg uncertainty, and is then consistent with ML-randomness of exit 
data from a site. Probability of 
existence of an atom in a site is given by a Langmuir-type isotherm and 
intrinsic exit properties are measured in V, as above (see \S 6  for the 
introduction of probability into $\mathbb{QKM}$ and \S 7 for final comments%
). By quantised complexity 
mechanics $\mathbb{QKM}$ we mean microscopic processes as described here. 
$A=(\ell _{N}/\lambda _{th})^{3}\gg 1$ will appear in state equations in 
$V$ for classical gas and $\lambda _{th}$ is reinterpreted for scales of 
pattern in $V$ in \S 4 below. 

Consider experimental apparatus such as that of \cite{Lipa1}, now with
collisionless gas. Lattice size in a copper vessel is $\sim 3\mathring{A}$
and estimate $\mathcal{M}_{tot}$. For $N\sim
10^{16}$ atoms and over a transit time $t_{b}\sim 10^{-4}s$ a flux of $\sim
10^{20}$\ atoms per second occurs at the wall. For wall of area $\pi
d^{2}=\allowbreak 3.8\times 10^{-3}m^{2}$ and given lattice size we find $%
\sim 0.25$ atoms at a site per transit time or one atom per site about every 
$4$ transit times of a free atom in $V$. We suggest that all atoms are 
decorrelated after two transits. At any moment there are three free
sites per occupied site allowing a simultaneous return flux to gas by the
heat bath. 

Algorithmic correlation by a macroscopic wall is broken and with $M=M(T,V)$ $%
D^{-}(T,V,N)$ \eqref{9} is now defined in $V$. Atoms disperse to
uniformly fill $V$ with thermalised energies born in quantised processes at
the real wall, thus accounting for $M$ much as did the energy eigenvalues of
the Schr\"{o}dinger problem for the smooth box. Emission energies are 
distinct and can be ordered from smallest to largest. 
By focusing only on the right side of \eqref{10}
and the ability of the gas in $V$ to do work, $D^{-}$ is an intrinsic
property of equilibrium gas under Fermi-Dirac combinatorics. By %
\eqref{10} a complete set of state equations $\mathfrak{E}^{-}$ follows for
fermions, given in \S 4. In the case of nonequilibrium gas (but with an
inbuilt \emph{relaxation process}) $K(\mathcal{L}_{i,t})$\ is always defined
but has no generally computable functional form. $T,P,c_{V},...$ are
strictly not defined. Detailed net disorder 
\begin{equation} \label{26}
D_{t}^{-}(\{\mathcal{L}_{Ms,t}\}_{s})/\ln 2=\sum_{s=up,down}K(\mathcal{L}%
_{M,t})-K(\mathcal{L}_{Ns,t})-K(\mathcal{L}_{M-Ns,t}),
\end{equation}%
always exists but no computable function can in general be associated. Each $%
K(\mathcal{L}_{is,t})$ is an absolute property of the data, only. Each is 
an \emph{intrinsic physical variable with units of
bits}, in equilibrium and nonequilibrium $\mathbb{TD}$. For a given gas we 
recall $\mathcal{L}_{t}$\ is instantaneous\emph{\ pattern in data} with
units of \emph{bits}, the complexity $K(\mathcal{L}_{t}\mathcal{)}$\ is
instantaneous \emph{quantity of pattern in data} with units of \emph{bits}
(for example $K(\mathcal{L}_{M})\approx 0$ for smooth box), and in
equilibrium \eqref{10} $k_{B}TK(\mathcal{L)}$\ is instantaneous\emph{\
energy of pattern in data }$\mathcal{L}$ with units of \emph{Joule-bits}. By 
$F=-k_{B}TD^{-}$, detailed net disorder $D^{-}$\ is a physical variable that
determines the quantity of Helmholtz free energy.\ All of this is objective 
and based on absolute properties of detailed data.

Lists of length $N\sim 10^{23}\sim 2^{76}$ with each datum to $\sim \log N$\
digits are beyond conceivable experimental detection but are strictly not to
be explicitly listed. We require only the existence of the new intrinsic
property of Kolmogorov complexity of an initial subsequence and its precise
interpretation as a thermodynamic variable. There is \emph{no feasibility
problem} that is a common justification for statistical mechanics \cite{B1}.

Measurements in thermodynamics need not approach $\sim \log N$\ digits. Via
Avogadro's number $N_{0}=6.0\times 10^{23}$, the gas constant and
Boltzmann's constant $k_{B}=\mathbf{R}/N_{0}=1.4\times 10^{-23}JK^{-1}$\ the
product $N_{0}k_{B}$\ of course reduces to a typical laboratory magnitude.
Tests of $\mathbb{QKM}$ ultimately lie with comparison of theoretical
prediction by gas/liquid laws and measurement to a few significant digits.\ 

\section{Re-interpretation of $\mathfrak{E}^{-}$, reversible processes}

Based on the assumption of ML-random equilibrium of collisionless $^{3}\text{%
He}$ gas data in $V$, with Kolmogorov complexity of initial subsequences we
gain considerable insight into inescapable effects of randomness alone. We
keep in mind that $\mathbb{QKM}$\ is not a statistical theory.
Fundamentally, $-k_{B}TD^{-}$ \eqref{10} is a new potential energy in
theoretical physics in that Kolmogorov complexity as quantity of pattern
does not appear in detailed $\mathbb{HM},\mathbb{SM},\mathbb{QSM},...$ nor
in $\mathbb{TD}$. Nowhere does an individual energy datum, nor mean value,
nor total energy, explicitly appear and at the fundamental level of
objective ML-randomness of data this changes the understanding of state
variables.

Define \emph{specific disorder }$\kappa (2A)$ for $^{3}\text{He}$\ before
taking limits by 
\begin{align}
D^{-}(T,V,N) &=N\kappa (2A)\text{, }\kappa (2A)=\ln (2A-1)+\Gamma ^{-}(2A),
\label{27} \\
\Gamma ^{-}(2A) &=2A\frac{\partial \kappa (2A)}{\partial (2A)}=-2A\ln
(1-(2A)^{-1}), \label{28} \\
A &=\gamma \frac{VT^{3/2}}{N}>1/2\text{, }\gamma =\left(\frac{2\pi mk_{B}}{h^{2}}%
\right)^{3/2}.  \label{29}
\end{align}%
These are objective and deterministic and ML-randomness and Kolmogorov
complexity play no further formulaic role. Equilibrium properties of the gas
now follow in analysis of extensive $D^{-}$\ and intensive $A$. We adjust
the statistical factor $2\pi $ in $\gamma $\ for objective probability in \S %
6. By working with $\log \binom{M}{N}$ on the right of \eqref{10} we
accept Pauli exclusion is effective and that the state equations are valid
for noninteracting structureless atoms into the fully quantum mechanical
regime of $\mathbb{QSM}$. In the case of structure additional disorders (eg
an orbital spin complexity for molecules) must be added in such a way that
extensivity of $D^{-}$ is maintained. The equations apply for noninteracting
mixtures, again with balance as in thermodynamics. Before taking limits of $%
A $, for $A\geq 1/2$, in $\mathbb{TD}$\ with $F=-k_{B}TN\kappa (A)$\ we
respectively find entropy, pressure and chemical potential given by 
\begin{align}
S &=-\frac{\partial F}{\partial T}=Nk_{B}[\kappa (2A)+\frac{3}{2}\Gamma
^{-}(2A)]=Nk_{B}[\kappa (2A)+e_{T}], \label{30} \\
P &=-\frac{\partial F}{\partial V}=\frac{Nk_{B}T}{V}\Gamma ^{-}(2A)=\frac{%
Nk_{B}T}{V}e_{V}, \label{31} \\
\mu &=\frac{\partial F}{\partial N}=-k_{B}T[\kappa (2A)-\Gamma
^{-}(2A)]=-k_{B}T[\kappa (2A)-e_{N}]=-k_{B}T\ln (2A-1). \label{32}
\end{align}%
\ Here we define \emph{thermal, volumetric and\ number elasticities }$%
e_{T},e_{V},e_{N}$, the change of specific disorder $\kappa (2A)$\ for
percentage change in $T,V,N$\ by 
\begin{equation}  \label{33}
e_{T}=T\frac{\partial \kappa }{\partial T}=\frac{3}{2}\Gamma ^{-}(2A)\text{, 
}e_{V}=V\frac{\partial \kappa }{\partial V}=\Gamma ^{-}(2A)\text{, }e_{N}=N%
\frac{\partial \kappa }{\partial N}=-\Gamma ^{-}(2A).
\end{equation}%
$\Gamma ^{-}(2A)$  (28) is elasticity with respect to $2A$\ and
carries the functional dependence of the elasticities. All elasticities are
intensive functions independent of the size of the system and apply in an
arbitrary subvolume of $V$. The fundamental role played by $\kappa $ and its
first derivatives is highlighted. They are closed algebraic formulae (no $%
\int $ signs) and have no adjustable free parameters. We refer to all such
properties derived from $D^{-}$ and first derivitives, as \emph{state
equations} denoted $\mathfrak{E}^{-}$.

Where $F$\ is energy of quantity of pattern in three incompressible lists in
gas, $S,P,\mu $ are formally potentials. Gas in $V$ is \emph{supported} in
equilibrium by these potentials against gas at the wall and by the copper
wall. They are not responses (compare $P$ with compressibility $%
-(1/V)(\partial V/\partial P$) of a nonequilibrium gas to a perturbation \cite%
{B1}. The condition for equilibrium is that $T,P,\mu $\ be equal in the two
populations \cite{B1}. Supports $P,\mu $ are consistent with the various
processes in the wall as associated with lists $\mathcal{L}_{i}$ in \S 3. It
is particularly expressive that $P$ is proportional to volumetric elasticity 
$V\partial \kappa /\partial V$ of gas and is independent of explicit
specific disorder $\kappa (2A)$. The chemical potential $\mu $ is finite but
relatively small ($\propto k_{B}T$) owing to on-going exchange of atoms with
the wall, the number elasticity $N\partial \kappa /\partial N$ cancels away
leaving only the $-\ln (2A-1)$ dependence of $\mu $; for $A>1$, $\mu <0$\
implies the gas exchanges atoms at the wall and with balancing equal $\mu $
in the wall is a support in this sense \cite{B1}. It depends only on that
part of specific disorder $\kappa $ independent of elasticity, that can be
traced back to quantities of pattern proportional to $N$ in lists $\mathcal{L%
}_{N}$ and $\mathcal{L}_{M-N}$ in $D^{-}$ in $V$. Thermodynamic entropy
appears as an extensive support $S$, it is dependent on size of the system
which may be very different in gas and wall (eg $N$ and $\mathcal{N}$) even
while $T$\ is the same in both gasses - thus $S$\ supports equality of $T$\
with gas and environment. $S/N$ is intensive and effective everywhere in $V$%
, it depends on full specific disorder $\kappa (2A)$ and on thermal
elasticity, also on all three lists in $D^{-}$ necessarily balanced by
processes in gas at the wall. From thermodynamics\ we find $S=(-F+U)/T$ and 
\emph{internal energy }%
\begin{equation} \label{34}
U(T,V,N)=Nk_{B}Te_{T}=\frac{3}{2}Nk_{B}T\Gamma ^{-}(2A).
\end{equation}%
$U$ is a partial support through $S$, representing thermal elasticity of the
gas. It is an extensive functional form independent of detailed
disorder. $U/N$ does not interpret as the average energy $E/N=(3/2)k_{B}T=%
\overline{\epsilon }$ of atoms of $\mathbb{QSM}$. We relate $E,U$\ in \S 6.

As potentials, the $S,P,\mu $ act simultaneously and independently to
mechanically support equilibrium gas against a detailed real wall. The
relation of support $S$ to von Neumann's subjective statistical entropy $%
S_{vN}$ and interpretation as a Shannon-type measure of information lacking
(or ignorance) is developed in \S 6 \cite{B1}. The ideal gas state equations 
$\mathfrak{E}_{0}$ follow for $A\gg 1$, $\Gamma ^{-}(2A)\approx 1$. At least
for $A\gg 1$ in collisionless gas,\ Sakur-Tetrode entropy $%
S_{ST}=S=Nk_{B}(\ln 2A+5/2)$ has a direct dependence on \emph{objective
detailed net disorder} $D^{-}$, in support of Clausius' intuition of his
entropy as disorder and admitting the arguments of $\mathbb{QSM}$.

Consider the $+--$\ signs in $D^{-}\simeq \sum\nolimits_{s}\ln \binom{M}{%
N_{s}}$\ \eqref{19}. It is known that complexity is subadditive, that
is, for energy lists concatenated such that $\mathcal{L}_{M}=\mathcal{L}%
_{M-N_{M}}\mathcal{L}_{N_{M}}$ 
\begin{equation}   \label{35}
K(\mathcal{L}_{M})\leq K(\mathcal{L}_{N_{M}})+K(\mathcal{L}_{M-N_{M}}|%
\mathcal{L}_{N_{M}})+O(1)
\end{equation}%
with equality (is additive) iff $\mathcal{L}_{M-N_{M}},\mathcal{L}_{N_{M}}$
are independent \cite{LV}. Applying (35) we write 
\begin{equation}   \label{36}
F=-kT_{B}D^{-}\simeq -k_{B}T(K_{N_{M}}-K_{N})=-k_{B}T(N\ln M-N\ln N)
\end{equation}%
where $K_{N_{M}}$\ is the complexity of occupied energy levels. Then $F$ is
the potential to do work by the energy of pattern of particles specifically,
but moderated by spatial relaxation processes. $F$\ remains extensive, $N\ln
M$\ with $M>N$ implies a well-defined list of intrinsic energies which is
physically significant compared to the independent list of locations of
atoms. Indeed it is in exact agreement with $\mathbb{QSM}$\ but is not
perfect (see \S 6). We refer to \eqref{36} as statistical $F_{stat}$.\ 

For completeness for $N$\ $^{4}\text{He}$\ bosons, state equations $%
\mathfrak{E}^{+} $ follow by substituting $\Gamma ^{-}(2A)\rightarrow \Gamma
^{+}(A)=A\ln (1+A^{-1})$, $\mu ^{-}(2A)\rightarrow \mu ^{+}(A)=-k_{B}T\ln
(A+1)$ for $A>0$.

Other equilibrium properties follow in pure $\mathbb{TD}$.\ Specific heat at
constant volume of the ideal fermion gas is 
\begin{equation}   \label{37}
c_{V}=\frac{\partial U}{\partial T}=\frac{3}{2}Nk_{B}\Gamma ^{-}(2A)+\frac{9%
}{4}Nk_{B}(\Gamma ^{-}(2A)-\frac{2A}{2A-1}).
\end{equation}%
As expected $c_{V}=(3/2)k_{B}N$, $A\gg 1$. The second term is thermal
elasticity of thermal elasticity of $\kappa (2A)$. An array of equilibrium
thermodynamic coefficients over second derivatives of $\kappa (T,V,N)$ is of
obvious interest but is not pursued here.

Of practical importance, the mathematical structure of pure thermodynamics
admits transformations among triplets of natural variables $T,V,N$, $T,V,\mu 
$, $T,P,N$ and $S,V,N$, $S,P,N$\ that are convenient for calculations. Given
(canonical free energy) 
\begin{equation}  \label{38}
F(T,V,N)=-k_{B}TD^{-}(T,V,N)
\end{equation}%
this is achieved by Legendre transformations to grand canonical and free
enthalpy energies 
\begin{equation}  \label{39}
A_{GC}(T,V,\mu )=F(T,V,N)-\mu N,
\end{equation}
and%
\begin{equation}   \label{40}
G(T,P,N)= F(T,V,N)+PV.
\end{equation}
$F(T,V,N)$ is the maximum potential to do work if $T=cnst$. The usefulness
of $A_{GC}(T,V,\mu ),G(T,P,N)$ lies respectively with potential to do work
if $T,\mu =cnst$ or $T,P=cnst$ and facilitates thermodynamic calculations.
Physically, the Legendre transformations respectively bring constant total
supports $-\mu N$, $PV$\ to $F$ so that the remaining terms in $F$\ are the
potentials to do the desired work. Free energies $A_{GC},G$\ cannot be
written in perfect $n\ln n$-terms, that is in terms of incompressible lists, 
are not formally associated with ML-randomness and are directly owing to
extensivness. The role of $F(T,V,N)$ and of $D^{-}(T,V,N)$\ in primitive
form is objective, unique and fundamental (so far for structureless
fermions). All this lies with thermo\emph{statics}. We have work done but
are yet to introduce the concept of heat.

Turn to thermodynamics. By instantaneous equilibrium we strictly mean (here
for fermions) that instantaneous \emph{incompressible} lists are
well-defined. Refer to \emph{nonautonomous} evolution under this condition
as a \emph{reversible} process. Langmuir-type phenomena in the wall are
completed and we consider the gas only.\ Denote $S=S^{\text{rev}}$. By
extensivity, associated with each of $F,A_{GC},G$ are differentials $%
dF,dA_{GC},dG$ in their respective natural variables \cite{B1}. A final
transformation 
\begin{equation}  \label{41}
U(S^{\text{rev}},V,N)=F(T,V,N)+TS^{\text{rev}}
\end{equation}%
introduces $S^{\text{rev}}$\ as an independent variable. From the Legendre
transformation \eqref{41}, in $S^{\text{rev}},V,N$ variables we find
the identity 
\begin{equation}  \label{42}
dU=PdV-\mu dN+TdS^{\text{rev}}.
\end{equation}%
With $U=F+TS^{\text{rev}}$ at the new equilibrium and before taking the
large-$A$ limit define 
\begin{align}
dU(S^{\text{rev}},V,N) &=-d(k_{B}T\sum\nolimits_{s}D^{-}(\{K(\mathcal{L}%
_{is})\}_{i})+TdS^{\text{rev}} \label{43} \\
&=PdV-\mu dN+TdS^{\text{rev}} \label{44} \\
&=-w+q. \label{45}
\end{align}%
In agreement with Clausius' definition of a reversible process, we can
identify work done on the gas by $w=-PdV+\mu dN$ and \emph{heat added} by $%
q=TdS^{\text{rev}}$ and now admit \emph{nonautonomous} changes $q,dV,dN$
such that equilibrium is continuously defined. While work and heat are
arbitrary external actions on the gas, $dU$\ is autonomous (spontaneous)
internal adjustment of partial support $U$ - for adequately slow such
actions $U=(3/2)Nk_{B}T\Gamma ^{-}(A)$ is continuously explicitly defined
and the change in $U$ at any moment is interpreted by the first equality
\eqref{43}. There is significant exchange of atoms with the changing
real wall, atoms are returned autonomously and one-by-one independently to $V
$ with relaxation processes owing to the wall. By \eqref{45} $dU$ is
consistent internal adjustment $d(k_{B}TN\kappa )$ of net disorder to
external work $w$ also to heat $q=TdS^{\text{rev}}$ added. By the latter we
see that heat added adjusts the mechanical \emph{support }%
\begin{equation}  \label{46}
dS^{\text{rev}}=k_{B}d(N\kappa +Ne_{T})
\end{equation}%
in reversible processes (note $q=TdS^{\text{rev}}\gtrless 0$).

The \emph{First Law for reversible processes} is then, as usual, 
\begin{equation}   \label{47}
dU=-w+q.
\end{equation}
It is \emph{conservation of energy-bits} in reversible processes in the form 
\begin{equation}    \label{48}
dU+w-q=0.
\end{equation}%
The First Law is \emph{not} related to detailed conservation of energy or of number
of particles. It is conservation of energies of pattern in thermodynamic
equilibrium, valid even while atoms come to rest in the wall.

The Carnot cycle and Second Law take the usual form in reversible
thermodynamics, the latter expressed by $\oint q/T=0$\ \cite{B1}.

Notably, thermal length $\lambda _{th}$\ \eqref{11} has not been
interpreted in $V$. At detailed level $\lambda _{th}$ relates to actions at 
marginal Heisenberg uncertainty. There is no local wave equation, interpretation 
as a de Broglie length is inappropriate, instead it is determined by a local 
scale of interaction in the wall governed by thermodynamic equilibrium. We 
distinguish classical gas in $V$\ where $M\propto T^{3/2}$
and $\lambda _{th}\propto T^{-1/2}$\ increases with decreasing $T$. Gas in $V
$ then decouples from scales in the wall. \ Equilibrium is defined by
quantities of patterns and we reinterpret $A=(\ell _{N}/\lambda _{th})^{3}$
as $A_{V}=V/\lambda _{V}^{3}$ where intensive $\lambda _{th}^{3}$ goes over
to extensive, size-dependent $\lambda _{V}^{3}=N\lambda _{th}^{3}$ defining
a volumetric scale of pattern as interactions at the wall map to the interior. 
We add that there is no nonlocal quantum entanglement for independent data, we 
consider equilibrium and so stimulated collective quantum fluctuations are 
excluded. The decoupling has no significant effect in collisionless
gas $A=6.1\times 10^{8}\gg 1$ where support $P=(Nk_{B}T/V)\Gamma
^{-}(2A_{V}) $, $\Gamma ^{-}(2A_{V})\approx 1$. We foresee that this will
change for quantum liquids where $A_{V}\lesssim 1$, $\Gamma ^{-}(2A_{V})\gg
1 $. Then $P/(Nk_{B}T/V)\gg 1$ hints at an unusual $T$-dependent degeneracy
pressure by Pauli exclusion at low density. $\mathfrak{E}^{-}$\ is quite
rich in implied phenomena for $A_{V}\lesssim 1$.

Reinterpretation and redefinition of strictly \emph{intensive} $A(T,V,N)$ is
the only freedom granted to definition of strictly intensive $\kappa (2A)$ in
fundamental $D^{-}=N\kappa (2A)$. This plays an important role in follow-up
theory of quantised boson liquids in $\mathbb{QKM}$.\ 

\section{Irreversibility and the Second Law in $V$}

We break equilibrium for irreversible thermodynamics in $V$. This requires
that we break incompressibility among three lists so that \eqref{8} is
not satisfied. Digits above the primitive $\log n_{i}$ level are
insignificant and to be physically significant, we turn to computably
undecidable integers $K(\mathcal{L}_{i})<\left\lfloor n_{i}\ln
n_{i}\right\rfloor $. Loss of homogeneity and/or isotropy is explicitly implied. 
For gas theory in $V$ relaxation demands Langmuir-type detailed interactions at 
the wall on timescales $\sim t_{b}$ consistent with reversible thermodynamics 
(\S 3). For fast time scales $\tau \ll t_{b}$ theories at macroscopic level must be 
reconsidered. A classical example is Joule expansion. Burst a balloon
suspended in a vacuum, as atoms approach the new real wall orientations of
collisionless trajectories have an intrinsic radial component, instantaneous
nearest neighbour displacements share a separating property which serves to
reduce always well-defined Kolmogorov complexty of the lists. Suppose two
balloons at different initial temperatures directly effecting anisotropy.
The perfect $D^{-}$ disorder is no longer physically significant in the face
of large-scale disturbance but a generally incomputable $D_{t}^{-}$ \eqref{26} %
remains well-defined, is physically most significant at
significant digit $<\log n_{i}$, is incomputable, and is to take place on
fast times $\tau \ll t_{b}$. Reversible thermodynamics is \emph{not} defined
on $\tau $. 

Formally \cite[pg215]{LV}, an infinite (instantaneous) data sequence $%
\mathcal{L}_{i}$ is \emph{not ML-random} iff initial subsequences satisfy 
\begin{equation}   \label{49}
\text{\emph{as }}n\rightarrow \infty \text{\emph{, }}\lim \sup
[n-K(x_{1:n})]=\infty .
\end{equation}%
Complexity of initial subsequences of a not ML-random sequence tend to lie
far below the incompressibility diagonal $K(x_{1:n})=n$. With \eqref{2}
there is then a typical \emph{complexity gap} between ML-random and
nonML-random gasses \cite{LV}. The latter are not in thermodynamic
equilibrium and the cool collisionless gas relaxes on timescales $\lesssim
2t_{b}=2b/v_{th}\lesssim 2\times 10^{-4}s\,$ as its atoms sample the real
wall. Runs of zeros and one's and repeated blocks of digits are
spontaneously removed from instantaneous data lists while tending to keep
low values $\sim \log MN$, but under the fundamental assumption ultimately
with sudden jump to a lower bound such that $K(\mathcal{L}_{it})\gtrsim l(%
\mathcal{L}_{it})= n_{i}\log n_{i}$ for strictly incompressible sequences
as the final such blocks are removed \cite{LV}. Granted relaxation we have
increase to the physically significant computable lower bound of
incompressibility but the integer functions are incomputable as equilibrium
sets in. It is striking that owing to the complexity gap, $K(x_{1:n}) \approx 0$ 
so that disorder is ignorable and deterministic physics applies. 

An \emph{adiathermal} reversible process ($q=0,dN=0$) in $T,V,N$-space say,
takes place in a surface of constant support $S^{\text{rev}%
}(T,V,N)=k_{B}N(\kappa (2A)+\frac{3}{2}\Gamma ^{-}(2A))=cnst$ \cite{Pipp}.
Intensive $A=cnst$\ so that $D^{-}=N\kappa (2A)=cnst$ also embeds
adiathermal reversible processes. Refer to a $D^{-}$ \emph{adiatherm}. Both
surfaces are also \emph{adiabats} ($q=0$) \cite{Pipp}. As above a
thermodynamic path in an adiatherm is reversible. Departure from adiathermal
change is of interest in relation to classic definitions of the Second Law
in pure thermodynamics, in particular Carath\'{e}odory associates the Second
Law: in a neighbourhood of a state there are states inaccessible by
adiathermal processes \cite{Cara},\cite{Pipp}. That is states not accessible
on any $A$- or $D^{-}(2A)$-adiatherm. A reversible
adiathermal process satisfies both $S^{\text{rev}}(2A)=cnst$\ and $%
D^{-}(2A)=cnst$ for any given $A(T,V,N)=cnst$. Work $-w=-PdV$ is done on the
gas as $T,V$\ are slowly varied. On the other hand if equilibrium is broken
support $S^{\text{rev}}$\ is not defined while well-defined time dependent
net disorder, a work-related function 
\begin{equation}  \label{50}
D_{t}^{-}/\ln 2=\sum\nolimits_{s}K(\mathcal{L}_{M{},t})-K(\mathcal{L}%
_{Ns,t})-K(\mathcal{L}_{M-Ns,t})>0
\end{equation}%
persists. \emph{Work} $-w=-P\delta V$ may be done \emph{on} the gas that
breaks equilibrium by sufficiently abrupt \emph{arbitrary} \emph{increases} $%
\delta V>0$\ in $V$ at $t=0$, while of course externally applied heat $q=0$.
The heat/particle bath is to remain defined by the new wall, \emph{processes}
in each term of \eqref{50} remain identified and the gas relaxes to a
new equilibrium with support $S(2A^{\prime })$\ and net disorder $%
D^{-}(2A^{\prime })$. Each term of \eqref{50} remains an absolute property of 
instantaneous data lists. This equilibrium with $A^{\prime }\neq A$ is not
accessible by an adiathermal process from a state in $A$. This
is not a reversible process and we refer to irreversible change $\Delta S^{%
\text{\text{irrev}}}$ to a new support $S(A^{\prime })$ (or for statistical
entropy where $\mathbb{QSM}$\ is well defined). By Carath\'{e}odory's
definition this expresses the Second Law. At the same time, by $\delta V>0$
and by the increasing lower bound of Kolmogorov complexity in \emph{%
spontaneous} relaxation, both $\Delta S^{\text{irrev}}>0$\ and $\Delta
D^{-}>0$ across the two equilibria. $D_{t=0}^{-}\simeq 0$ immediately (ie
drops), a well-defined but incomputable relaxation $D_{t}^{-}$ takes place
with, as above, a step-like jump across the complexity gap to the new
equilibrium $D^{-}(A^{\prime })$. Fundamental Fermi-Dirac combinatorics is
not effective during irreversible change because new detailed sites on the
wall are defined but are not immediately effective, detailed Pauli exclusion
remains effective. Joule expansion with $w=0$ and Joule-Thomson throttled
expansion $\Delta U=-P\delta V$\ are classic examples of irreversible
processes and where arbitrary work is exhibited \cite{B1}. Clausius'
conception of irreversible processes together with his intuition that
``disorder'' increases are satisfied in respect of $S^{\text{irrev}}$\ and $%
D^{-}$. This is now given an objective footing, at least for ideal
noninteracting gasses.

The First Law holds across the two equilibria because the new state can be
reached by a reversible process, but is not defined during relaxation,
because partial support $U$ is defined for equilibria. Energy per atom is
not conserved in general, given the real wall.

The formal equivalence of Carath\'{e}odory's definition of the Second Law
with the definitions of Clausius and Kelvin is noted \cite{B1}. In
coarse-grained $\mathbb{QSM}$, von Neumann's statistical entropy in the
ideal gas limit (Sackur-Tetrode entropy) is a form of the Second Law 
\begin{equation}  \label{51}
\Delta S_{vN}=\Delta k_{B}N(\kappa (2A)+3/2)=\Delta S^{\text{rev}%
}=k_{B}\Delta D^{-}>0,
\end{equation}%
justifying a statistical Second Law. As a support we here find that equally 
\begin{equation}  \label{52}
\Delta \mu =-\Delta k_{B}T(\kappa (2A)-1)=-k_{B}T\Delta D^{-}<0
\end{equation}%
for adiathermal processes suggestive of Boltzmann's $H$-function and can
apparently serve as a Second Law as well, granted particle exchange in the
wall. We naturally favour direct characterisation of the Second Law by an
unambiguous, objective, increasing disorder in irreversible processes with
relaxation, that is by 
\begin{equation} \label{53}
\Delta D^{-}>0
\end{equation}%
for adiathermal processes across two equilibria. It is significant that the
support $S^{\text{rev}}$ diverges as $A\rightarrow 1/2^{+}$ where net
disorder $D^{-}=0$, so that reversible (algorithmic) $S^{\text{rev}}$ is 
\emph{not} always a measure of disorder. The collapse of $D^{-}$\ is
physically meaningful, because $M-N_{s}=0$\ and independence of lists
collapses.

The \emph{necessarily detailed} processes in the real wall are
nondeterministic in that the history of each emitted atom is lost and Born
probability of energy and orientation applies on emission. There is then no
Thomson-Loschmidt paradox that macroscopically irreversible processes are
owing to reversible microscopic dynamics \cite{B1}. There is not a new
paradox that reversible processes emerge from irreversible detailed
mechanics (granted relatively fast relaxation). The Zermelo-Poincar\'{e}
recurrence paradox does not arise owing to interrupted trajectories at the
wall \cite{B1}. However we are not satisfied that $D_{t}^{-}$ \eqref{50}
admits a ``strong'' Second Law of always increasing value. Spontaneous
relaxation is necessary and is built into quantum Fermi-Dirac combinatorics.
Conversely it is only to be applied where there is relaxation to
equilibrium. In $\mathbb{QKM}$ relaxation owing to objective spontaneous
ML-randomisation (here by the wall) is yet another possible statement of the
Second Law. Clausius cosmology (heat death of the universe owing to Second
Law of thermodynamics) is not obvious. In particular, accelerating expansion
of a flat cosmos is observed \cite{galax}. Some additional large scale
potential is implied and future Kolmogorov complexities of intrinsic lists
in the face of local gravitational attraction are not clear.\ 

\section{Probability on the left and of $\mathbb{QSM}$}

The left equality of \eqref{10} is mathematically valid and it follows
that the left hand term is physical, consistently with ML-randomness of the
lists. The original motivation behind mathematical theory of ML-randomness
and Kolmogorov complexity lay with rigorous foundation for probability \cite%
{LV}. Central to this was Martin-L\"{o}f's definition of randomness and
concepts of tests of randomness \cite{ML}. We now have $\sum\nolimits_{s}%
\binom{M}{N_{s}}$, ML-randomness implied, as necessary and sufficient in its
significant extensive part for detailed equilibrium of fermion gas. We turn
away from physics of hidden processes and seek insight to $%
FD_{M,N}=1/\sum\nolimits_{s}\binom{M}{N_{s}}$, a recursive probability
distribution. With regard to probability we state only needed technical
results, taken from \cite[sections 4.2,4.3]{LV}.

Directly from \cite[Example 4.3.9]{LV} a \emph{universal sum test} with
respect to the Fermi-Dirac distribution is 
\begin{equation}  \label{54}
\forall x,M,N\in \mathbb{N}\text{, }\kappa _{0}(x|FD_{MN})=\log\left (\frac{%
\mathbf{m}(x)}{FD_{MN}(x)}\right).
\end{equation}%
Here $\kappa _{0}(x|FD_{MN})$ is \emph{randomness deficiency} of a finite
binary sequence $x\in \mathbb{N}$\ with respect to Fermi-Dirac distribution. 
$FD_{MN}$\ is necessarily an example of a recursive probability
distribution, in general $P(x)$ say. There are three central features in the
relation of ML-randomness to probability. First, for $x\in \mathbb{N}$, $%
\mathbf{m}(x)$ is a universal, enumerable discrete semimeasure. A discrete
semimeasure is a function $P(x):\mathbb{N}\rightarrow \mathbb{R}$ such that $%
\sum\nolimits_{x\in \mathbb{N}}P(x)\leq 1$ and is a probability measure if
equality holds (there is no difference between a discrete measure and a
probability distribution over sample space $\mathbb{N}$ \cite{LV}). In a
class $\mathcal{M}$\ of semimeasures ($\mathcal{M}$ is not Langmuir adhesion
sites), enumerable means that there is an indexed sequence of recursive
semimeasures $P_{1}(x),P_{2}(x),...\in \mathcal{M}$ such that $%
P_{i}(x)=P(i,x)$ for every $i,x\in \mathbb{N}$. A semimeasure $P_{0}$ is 
\emph{universal} (or maximal) for $\mathcal{M}$ if $P_{0}\in \mathcal{M}$\
and for recursive $P\in \mathcal{M}$\ there exists a constant $c_{P}$\ such
that for all $x\in \mathbb{N}$, $c_{P}P_{0}(x)\geq P(x)$ and where $c_{P}$\
may depend on $P$ but not on $x$ (for $c_{P}$ see below). $P_{0}$\ is said
to dominate recursive semimeasures in $\mathcal{M}$. Finally, a $P_{0}(x)$
exists denoted $\mathbf{m}(x)$, it is not recursive and $\sum\nolimits_{x}%
\mathbf{m}(x)\leq 1$ \cite[Theorem 4.3.1, Lemma 4.3.2]{LV}. That $\mathbf{m}%
(x)$ dominates all computable semimeasures gives insight into the universal
sum test \eqref{54} and by the recursive $P(i,x)$ to Martin-Lof's very
strong definition of randomess, that it satisfies all computable tests of
randomness with respect to $FD_{MN}$. Further tests for example might
include simple counts of zeros and ones and examination for convergence of
their ratio. Second, it can be shown that there is a \emph{universal} \emph{%
prior probability }%
\begin{equation} \label{55}
Q_{U}(x)=\sum\nolimits_{U(p)=x}2^{-l(p)}\text{, }\sum\nolimits_{p}2^{-l(p)}%
\leq 1,
\end{equation}%
defined for all algorithms $p$ of length $l(p)$\ that halt on the prefix $p$%
\ over \emph{arbitrarily long} sequences of flips of a fair coin (reference
prefix machine $U$ \cite[Definition 4.3.3]{LV}). Then $2^{-l(p)}$ is uniform
probability of throwing $x$. Third, \emph{algorithmic probability} is
defined by $R(x)=2^{-K(x)}$, so a high probability indicates a simple
regular object (eg data list, $K(x)\simeq 0$) and in the present application
a high probability of a distinct physical phenomenon (eg smooth box is not
mixing). These are defined by short sequences, rare among all sequences. Low
probability $1/2^{l(x)}$ indicates effectively always ML-random coin tosses
in uniform probability with no computational means to characterise a
physical difference between any two. Algorithmic probability enacts Occam's
razor, that one chooses the simplest description of $x$ - universal $Q_{U}(x)
$\ dominates all possible recursive prior probabilities so is most likely to
be true \cite[Chapter 4]{LV}. The Coding Theorem states that for every
sequence $x\in \mathbb{N}$ 
\begin{equation}  \label{56}
-\log \mathbf{m}(x)=-\log Q_{U}(x)=K(x),
\end{equation}%
with equality up to additive constant $c$ \cite[Theorem 4.3.3]{LV}. The
Conditional Coding Theorem\ is 
\begin{equation}  \label{57}
-\log \mathbf{m}(x|y)=-\log Q_{U}(x|y)=K(x|y)+O(1).
\end{equation}%
Like $K(x),K(x|y)$, $\mathbf{m}(x),\mathbf{m}(x|y)$, $Q_{U}(x),Q_{U}(x|y)$\
are not recursive. Finally \cite[Theorem 4.3.5]{LV}, $P$ is a recursive
probability distribution, (57) admits $y=P(x)$ and 
\begin{align}
\forall x &\in \mathbb{N}\text{, }\kappa _{0}(x|P(x))=\log (\mathbf{m}%
(x)/P(x)) \\
&=-K(x)-\log P(x)+O(1)
\end{align}%
is a \emph{universal} sum $P$-\emph{test} for randomness (for any $x\in 
\mathbb{N}$\ and thus $K(x)$\ whether incompressible or not) and $\kappa
_{0}(x|P(x))$ is randomness deficiency. This name arises because $\kappa
_{0}(x|P(x))=O(1)$ if and only if $x$ is ML-random, that is if $K(x)=-\log
P(x)+O(1)$.

Directly from \cite[Example 4.3.9]{LV}, \eqref{54} follows with
randomness deficiency $\kappa _{0}(x|FD_{MN})=O(1)$ for $P(x)=FD_{MN}(x)$
and arbitrarily long sequences $x\in \mathbb{N}$, assuming ML-randomness of
lists in collisionless fermionic gas. We find \emph{algorithmic probability} 
\begin{equation}  \label{60}
FD_{MN}=2^{-\sum\nolimits_{s}\log \binom{M}{N_{s}}}\lesssim 2^{-D^{-}/\ln
2}=z_{C}^{-1}
\end{equation}%
as on the left side of \eqref{10}. Thus $z_{C}$\ is the algorithmic
partition function appropriate to Helmholtz free energy $F$ and energy of
pattern on the right. All this is taken from \cite{LV}. However we now acknowledge that a sequence $x\in 
\mathbb{N}$ \emph{is conditional on relaxation}.

Define $\mathcal{L}_{M}^{\ast }:l(\mathcal{L}_{M}^{\ast })=l_{M}^{\ast
}=M\log M+(1/2)\log M$ which includes the $\log \sqrt{M}$ nonextensive term
in \eqref{6}. As usual $l(\mathcal{L}_{M})=M\log M$ is the physically
significant primitive list for well-defined $M$ so that a few $\log \sqrt{M}%
=\allowbreak 29$ digits are concatenated to define $l(\mathcal{L}_{M}^{\ast
})$ in our gas. Similarly for $\mathcal{L}_{N}^{\ast },\mathcal{L}%
_{M-N}^{\ast }$. We are interested in a class of semimeasures consistent
with reversibly concatenated instantaneous lists $x=\mathcal{L}^{\ast }%
\mathcal{=L}_{M}^{\ast }\mathcal{L}_{N}^{\ast }\mathcal{L}_{M-N}^{\ast }$ of
constant complexity in thermodynamic equilibrium, physically significant and
consistent with Pauli exclusion, homogeneity, isotropy. We simplify by
recalling the subadditive property \eqref{35} so that for the two lists
of energies we may define $K(\mathcal{L}_{M}^{\ast })-K(\mathcal{L}%
_{M-N}^{\ast })=K(\mathcal{L}_{N_{M}}^{\ast })$ where $\mathcal{L}_{M}^{\ast
}=\mathcal{L}_{N_{M}}^{\ast }\mathcal{L}_{M-N}^{\ast }$, thus $x=\mathcal{L}%
^{\ast }\mathcal{=L}_{N_{M}}^{\ast }\mathcal{L}_{N}^{\ast }$ where $\mathcal{%
L}_{N_{M}}^{\ast }=\mathcal{L}_{N_{\text(up}}^{\ast }\mathcal{L}_{N_{\text%
(down}}^{\ast }$\ is the list of occupied energy levels. The following
applies also in $\mathbb{QSM}$, relaxation understood, in equilibrium gas.
For all $M,N=2N_{s}\in \mathbb{N}$ 
\begin{align}
K(x =\mathcal{L}^{\ast })&=-\log FD_{M,N}+O(1) \label{61} \\
&=K(\mathcal{L}_{N_{M}}^{\ast })-K(\mathcal{L}_{N}^{\ast })+O(1). \label{62}
\end{align}%
Both physically significant initial subsequences of $FD_{MN}$ are recursive.
Now $K(\mathcal{L}_{i}^{\ast })=-\log (2^{-l_{i}^{\ast }})$ and $L_{i}^{\ast
}=2^{-l_{i}^{\ast }}$ is \emph{discrete Lebesgue measure}, the uniform
probability distribution over the space of all possible initial subsequences
for each $i$. Turn to randomness deficiency in the sum test for such
measures \cite[Example 4.3.6]{LV}. Processes are restricted to a subset $%
Y_{i}^{\ast }=\{0,1\}^{l^{\ast }}\subset \mathbb{N}$\ and $\forall y\in
Y_{i}^{\ast },\kappa _{0}(y|L_{i}^{\ast })=l_{i}^{\ast }-K(y|l_{i}^{\ast
})-O(1)$. For necessarily ML-random sequences $\mathcal{L}_{i}^{\ast }$, $%
K(y|l_{i}^{\ast })=n_{i}\log n_{i}+\log \sqrt{n_{i}\log n_{i}}=l_{i}^{\ast }$
so that each term passes a randomness deficiency test as does $K(x=\mathcal{L%
}^{\ast })$ for $FD_{MN}$ \eqref{62}, by 
\begin{align}
\kappa _{0}(x|FD_{MN}(\mathcal{L}^{\ast })) &=\log (\mathbf{m}(\mathcal{L}%
^{\ast })/FD_{MN}(\mathcal{L}^{\ast })) \label{63} \\
&=K(\mathcal{L}^{\ast })-\sum\nolimits_{s}\log FD_{MN}(\mathcal{L}^{\ast
})=O(1). \label{64}
\end{align}%
By \eqref{63} and the conditional coding theorem a universal prior
probability exists for each list, that is 
\begin{equation}  \label{65}
Q_{U}(\mathcal{L}_{i}^{\ast })=2^{-K(\mathcal{L}_{i}^{\ast })}.
\end{equation}%
For equilibrium gases extensive $D^{-}(\mathcal{L})$ is necessary and
sufficient and the term $\delta ^{-}(M,N)=(1/2)\log (M/N(M-N))$ lies with
physically insignificant digits. For incompressible lists we find randomness
deficiency\emph{\ }%
\begin{equation}   /label{66}
\kappa _{0}(\mathcal{L}|2^{-D^{-}})=-D^{-}(\mathcal{L})-\log FD_{MN}(%
\mathcal{L}^{\ast })=\delta ^{-}\approx \log N^{1/2}=30
\end{equation}%
in our application. Admitting this deficiency in randomness is formally a
weak \emph{regularising} effect (it admits equilibrium). It is known that
the wedge of integer functions defined by 
\begin{equation}   \label{67}
n+O(1)\leq K(x_{1:n})\leq n+K(n)+O(1)
\end{equation}%
contains all and only ML-random sequences \cite{LV}. We note these
complexity oscillations take place at less significant digits than the
extensive primitive digits and find randomness deficiencies $\kappa _{0}(%
\mathcal{L}|2^{-K(wedge)})\lesssim 40$ for $N\ $and $\lesssim 61$ for $M$. 
Admitting these deficiencies is similarly weak regularisation, and it is
these that set benchmarks for admitting equilibrium thermodynamics. To do
work it is critical that primitive extensive lists are effectively always
incompressible and this is not affected. $D^{-}$ is maximal and thermodynamic work 
potential $F$ is minimal as required for thermodynamic equilibrium.

Return to the constant $c_{P}$\ such that $\forall x,c_{P}\mathbf{m}(x)\geq
P(x)$\ for universal $\mathbf{m}$. It is known via Markov's Inequality that $%
c_{P}=2^{K(P)}$ and that the probability that $c_{P}^{-1}\mathbf{m}(x)\leq
P(x)\leq c_{P}\mathbf{m}(x)$ is $1-1/c_{P}$. $K(P)=O(1)$ here is the
description length of discrete Lebesgue measure $L_{i}^{\ast }$ requiring
only its index not the lists \cite[Example 4.3.10]{LV}. Then for $%
L_{i}^{\ast }$ each is in probability the closest possible recursive
estimate for its nonrecursive appropriate $\mathbf{m}(\cdot )$. It will be
noticed that $\kappa _{0}(x|\mathbf{m})=O(1)$ which declares all $x$ to be
ML-random. The physical implication is that maintenance of and relaxation to
thermodynamic equilibrium under Pauli exclusion is close to most efficient
possible. This is not apparent from $FD_{MN}(x)$ as used in $\mathbb{QSM}$.

The physically significant algorithmic partition function and free energy
are 
\begin{align}
z_{C}(T,V,N) &=2^{D^{-}/\ln 2}=e^{N\kappa (2A)}, \label{68} \\
F(T,V,N) &=-k_{B}TN\kappa (2A). \label{69}
\end{align}%
With supports $U=(3/2)(Nk_{B}T/V)\Gamma ^{-}(2M/N)$ and chemical potential $%
\mu =-k_{B}T\ln (2M/N-1)$ and with algorithmic probability $%
e^{-D^{-}}=(e^{-\kappa ^{-}})^{N}$ we define 
\begin{equation}   \label{70}
g^{-}=e^{-\kappa ^{-}}=\frac{1}{e^{(U/N-\mu )/k_{B}T}}\text{, }%
0<g_{MN}^{-}\leq 1.
\end{equation}%
In $V$ we define $\epsilon ^{-}/k_{B}T=U(2A)/(3/2)Nk_{B}T=\Gamma ^{-}(2A)$,
whence 
\begin{equation}    \label{71}
g^{-}(2A)=\frac{1}{e^{(\epsilon ^{-}-\mu ^{-})/k_{B}T}}=\frac{1}{e^{(\Gamma
^{-}(2A)+\ln (2A-1))}}\text{, \emph{for} }\frac{1}{2}<A<\infty .
\end{equation}%
Then $g^{-}(2A)$\ takes the place in $\mathbb{\mathbb{QKM}}$ which the
expectation value of an energy level $\epsilon $ in \emph{Fermi-Dirac
statistics }%
\begin{equation}   \label{72}
f^{-}(\epsilon )=\frac{1}{e^{(\epsilon -\mu )/k_{B}T}+1}
\end{equation}%
takes in $\mathbb{QSM}$ \cite{B2}. Quantum phenomena occur at a Fermi
surface $\epsilon =\mu $, here where $A=1/2$ (but not appropriate for
collisionless gas). In $\mathbb{QKM}$ energy levels are independent and have
uniform distribution $2^{-M\log M}$ at fixed $T,V,N$. Here $g^{-}$ is
directly maximal algorithmic probability of an instantaneous list under $%
\kappa ^{-}(M,N)=\log (FD_{MN})/N$. Equilibrium is determined with exchange
of atoms in the wall, the relation between $\epsilon ,\mu $ is dependent
only on $A=M/N$. The two distributions closely agree for $A\gg 1$,\ $\Gamma
^{-}(2A)\approx 1$ but are qualitatively different.

All the above carries through for Bose-Einstein combinatorics $%
P(x|M,N)=BE_{MN}=\binom{M+N}{N}^{-1}$ \cite[Example 4.3.9, also pg261]{LV}.
For completeness, for bosonic $^{4}\text{He}$ and $BE_{MN}(x)$ from \eqref{21} %
$\Gamma ^{+}(A)=A\ln (1+A^{-1})$, $\mu ^{+}=-k_{B}T\ln (A+1)$ for 
$A>0$\ so that Bose-Einstein statistics $g^{+}(A)=1/e^{(\epsilon ^{+}-\mu
^{+})/k_{B}T}$ follows \cite{B2}. Sequence $x\in \mathbb{N}$\ is again
conditional on relaxation.

Well-defined lists of distinct data can be ordered from smallest to largest.
Rather than spreads and means of statistics we emphasise notions of
ML-randomness. We have that incompressible lists satisfy effectively all
tests of randomness including that initial subsequences have roughly equal
numbers of zeros and ones. Energy levels are naturally increasing lists,
packed with zeros at lower values ones at higher values. If $m=\left\lfloor
M\log M\right\rfloor $\ digits,\ $m$ even say, define a regrouped sequence $%
\omega $ of $2m$\ integers in $\sqrt{m}$ digits where $\{\omega
_{j}\}=\{00...0,00...1,...,01...1\}$\ on the interval $[0,1/2)$, similarly
for the interval $[1/2,1)$ for a total of $2^{m}$ digits. In groups of $%
\sqrt{m}$ integers it will be found that effectively always (a fraction $%
1-2^{-m}$ \eqref{3}) the count of nearly balanced zeros and ones is
maximised in the intervals just before and just after $1/2$, falling off as
zeros are added to the left, ones to the right. Localisation of the
distribution of such balanced states is very strong for $M=6.1\ast 10^{24}$, 
$\sqrt{M}=2.\,\allowbreak 5\times 10^{12}$. ML-random energy levels are a
natural grouping around mid-interval $(m/2)(1\pm 1/(2\sqrt{m}))$.
Incompressible lists of distinct $N=N_{M}<M$ occupied levels and $M-N_{M}$
empty levels can be arranged in increasing order, have grouping $\pm \sqrt{%
m_{N}}/2=\pm \sqrt{N_{M}\log N_{M}}/2,\pm \sqrt{m_{M-N}}/2=\pm \sqrt{%
(M-N_{M})\log (M-N_{M})}/2$. For $A=M/N\gg 1/2$, ratios $\sqrt{m}/\sqrt{%
m_{M-N}}\sim 1$ so that empty levels cover the full range of $\mathcal{L}_{M}
$, the ratios $\sqrt{m_{M-N}}/\sqrt{m_{N}}\sim \sqrt{2A}=\allowbreak
1.4\times 10^{4}$ so occupied levels are peaked relative to empty levels. Where 
description length is primitive $nlogn$ form for nonrandom data, it is acceptable 
in equilibrium. The fraction of compressible lists with shorter description length 
in the peak is effectively zero by \eqref{3} and equal a priori probability applies
with ignorable risk. Order ML-random (distinct) nearest neighbour locations,
also orientations, from smallest to largest. Similar peaks exist for
balanced zeros and ones about $\ell _{N}=(V/N)^{1/3}$ for effectively
uniformly distributed locations by nearest neighbour displacements, and for
orientations $\mathcal{L}_{\theta }^{\prime }$. The effect of the real wall
as heat/particle bath is to simultaneously maintain the number of atoms in
peaks. $A^{-1}=N/M$ is a measure of typical atoms per energy level in the
peak, not including levels near ground or maximum. \emph{Relaxation driven
by the real wall spontaneously brings any disturbance of the gas back to
ML-random levels and incompressible lists}.\ The fast fall-off from the peak
of occupied states to unoccupied state is consistent with rapid switching
across the complexity gap between \eqref{2}\eqref{49}, of ML-random
and nonrandom gases. Together it is these lists that define uniform
semimeasures $2^{-K(\mathcal{L}_{i})}$ such that equal a priori probability
is justified in $\mathbb{QSM}$.

To this point we have been working with $\lambda _{th}$ as defined in $%
\mathbb{QSM}$ \eqref{11}. In $\mathbb{QKM}$\ it is of interest to
standardise on Weyl's objective expression for the number of levels spanned
to some $\epsilon _{\max }$ under a one-particle Schr\"{o}dinger problem,
where for macroscopic $\partial V$ distinct levels discretise classical
phase space in units of $h^{3}$\ which interprets $M_{Weyl}=\frac{4\pi }{3}%
V(2m\epsilon _{\max }/h^{2})^{3/2}$ \cite{BandB}. \emph{This rule does not
apply here} but for given span $M$ it does serve as a mid-point for spread of independent
levels in $V$. Setting $M_{Weyl}=M_{\mathbb{QSM}}$\ we find $\epsilon _{\max
}=0.81\overline{\epsilon }$, for Gaussian distributions the mode and mean
satisfy $\epsilon _{\text{\emph{mode}}}=0.82\overline{\epsilon }$ and
pursuing objective physics\ in $\mathbb{QKM}$ we interpret energies at the
peak rather than standard deviation of energies about a mean as in $\mathbb{%
QSM}$. The separation in statistics is consistent with levels unbounded
above. Finally define $M_{Weyl}=V/\lambda _{\mathbb{QKM}}^{3}\ $so that 
\begin{equation}    \label{73}
\lambda _{\mathbb{QKM}}(\text{{\small mode}})=\lambda _{th}(\text{{\small %
mean}}).
\end{equation}%
The interpretations $\lambda _{V}^{3}=N\lambda _{th}^{3}$ in $V$ and $%
\lambda _{th}$ itself as scale of a Langmuir-type site are unchanged.

$\mathbb{QSM}$ is a competing theory but is qualitatively different. 
By Pauli exclusion and discrete energies, a fermionic sample space of
cardinality $\sum\nolimits_{s}\binom{M}{N_{s}}$ is common to both $\mathbb{%
QSM}$ and $\mathbb{QKM}$\ but for different purposes. $\mathbb{QSM}$ differs
in two ways, first that following Clausius and Boltzmann thermodynamic
entropy rather than thermodynamic free energy $F=-k_{B}TD^{-}$\ associates
``disorder''. Second, to get to the increasing Second Law it is necessary to
coarse grain, to introduce ignorance, as in construction of density
matrices. The quantised theory originates with von Neumann \cite{B2}. Via
pure thermodynamics the outcome is Fermi-Dirac statistics $f^{-}(\epsilon )$
\eqref{70}. The extraordinary success of $\mathbb{QSM}$ is its own
justification (perhaps strengthened now with evidence for mixing and equal a
priori probability). Our present theory scratches the surface. We presume
only to question the use of von Neumann's statistical entropy $S_{vN}$
rather than $F $\ as above and now reframe $\mathbb{QKM}$\ in such terms.
From $\mathbb{TD}$, $TS_{\mathbb{QKM}}=U-F$, $U=\frac{3}{2}Nk_{B}\Gamma ^{-}$%
\ and $\Gamma ^{-}=-(2M/N)\ln (1-N/2M)=1+N/4M+O(A^{-2})$ and we have 
\begin{equation}  \label{74}
S_{\mathbb{QKM}}/k_{B}\ln 2=K(\mathcal{L}_{M})-K(\mathcal{L}_{N})-K(\mathcal{%
L}_{M-N})+\frac{3}{2}N(1+O(A^{-1}).
\end{equation}%
This is extensive and at least for thermostatics for given fixed $N$ (closed
systems), 
\begin{equation}   \label{75}
\Delta S_{\mathbb{QKM}}/k_{B}\ln 2=\Delta D^{-}
\end{equation}%
across two equilibria is rigorously in agreement with $\mathbb{QKM}$, that
is, change of algorithmic entropy is justifiably a good interpretation of
change of detailed net disorder. Both are automatically maximised, formal
justification for the maximum entropy principle in $\mathbb{QSM}$.

$\mathbb{QSM}$\ takes explicit account of particle energy only. In $\mathbb{%
QKM}$ if inhomogeneity is possible the term $K(\mathcal{L}_{N})$ in $D^{-}$\
collapses. From randomness deficiency \eqref{63} we find 
\begin{equation}   \label{76}
\sum\nolimits_{i=M,M-N/2}\kappa _{0}(y|L(\mathcal{L}_{i}^{\ast }))=K(%
\mathcal{L}_{N})\lesssim N\log N.
\end{equation}%
Interpreting the right hand side as risk of inhomogeneity before running an
experiment, the risk is a significant increase on the baselines following
\eqref{65}. For the smooth box or generally when energies are
algorithmically correlated by the wall the risk is $K(\mathcal{L}%
_{M})\lesssim M\log M$ if no relaxation processes is identified. Among
general applications of Kolmogorov complexity theory is risk (and reward)
taken by a gambler in games of chance \cite{LV}. Regarding a theoretician as
a gambler, risk is quantified in $\mathbb{QKM}$. Risk in foundations of
thermodynamics where homogeneity and isotropy are ignored recalls above
doubts of an always true Second Law where relaxation is in doubt. The risk
of assuming thermodynamic variables in $\Lambda CDM$ such as density and
pressure on cosmic scales without addressing uniformity of matter is $O(N)$\
at least \cite{galax}. Relaxation of an expanding cosmos to equilibrium is
not clear.

Restricting $\mathbb{QSM}$ to observably relaxing gasses as we do in $%
\mathbb{QKM}$, both are good theories on some domain of $T,V,N$ say, in a
Bayesian sense of so far so good. Yet concerning this, given equilibrium
across a history of repeated experiments and applications, in $\mathbb{QKM}$%
\ for ML-random lists the objective prior probability $Q_{U(p)=x}$ \eqref{55} %
maximises all computable probabilities over the history, gives
the absolutely fastest convergence in Bayesian sense (Bayes-Solomonov
inference \cite{BaySol},\cite{LV}), also, is most likely in objective
probability to be applicable. Consider again the $+--$\ signs in $%
D^{-}\simeq \sum\nolimits_{s}\ln \binom{M}{N_{s}}$\ \eqref{19}. Then $%
2^{-D^{-}}\simeq 2^{-K(N/2)}2^{-K(M-N/2)}/2^{-K(M)}=R(N/2)R(M|N/2)/R(M)$
hinting at algorithmic probability $2^{-D^{-}}=R(N/2|M)$ as physical
realisation of Bayes-Solomonov inference that independent particle data is
owing to $M$ independent energy levels. Furthermore, the intrinsic
relaxation process is fastest possible and maximally efficient.

For equilibria\ interpretation $U=k_{B}Ne_{T}$ of internal energy as a
partial support is consistent with the assumption of classical atoms in $V$
and is then naturally compared to interpretation $E$ as total kinetic energy
in $\mathbb{QSM}$. Support $U$\ is formally a potential energy owing to the
everywhere location, orientation and energy of atoms, which for elemental
volumes is foreseen as a detailed foundation for equilibrium in continuum
mechanics. Maxwell's explanation of $E$ in kinetic theory as total
translational energy is not obviously a support but is consistent with
interpretation of $P$ as a net effect of ballistic bombardment of the wall
by atoms which is force per unit area on the wall, so $E=3P/2$\ is
indirectly a support for equilibrium. It is of interest that Maxwell fully
appreciated an assumption of uniformity (1860). $\mathbb{QKM}$ is built on
an intrinsic property, quantity of pattern of data - a piston may be
supported by an equilibrium gas with incompressible $K(\mathcal{L})$, it may
be supported by the ballistic impact of a directed gas jet with compressible
(ie ignorable) $K(\mathcal{L})$. The notion of quantity $K(x)$ of pattern
and of potential energy of pattern is appropriate, as is identification of a
new physical variable in a new physical theory. 

Finally, energies of pattern $k_{B}TK(\mathcal{L}_{it})$ have units of 
Joule-bits. Use of ``bits'' hints at ``information'' but as for Shannon entropy 
in $\mathbb{QSM}$, ``information'' is ``message conducted from an ensemble 
thereof'' \cite{LV}\cite{B1}. $D^{-}$ \eqref{8} is a Kolmogorov 
complexity and may be read as a quantity of pattern in a gas that is \emph{%
not} that of $N$ locations and \emph{not} that of $M-N$ unoccupied energy
states, which admits the interpretation of entropy as a measure of
``ignorance'', as in $\mathbb{\mathbb{\mathbb{\mathbb{\mathbb{QSM}}}}}$. On 
the right of \eqref{9}, extensive $D^{-}$ in terms of $\{K(\mathcal{L}%
_{i})\}_{i}$\ formally carries the existence of algorithms (implied but not 
realised) to \emph{exactly} write detailed data to a precision necessary and 
sufficient for the same purposes as $\mathbb{QSM}$. The ``unknown'' part that 
is taken as disorder in $\mathbb{QSM}$ is exactly ``known''. $D^{-}$ itself is 
known where $F$\ is known in applied thermodynamics. Hidden mechanics has 
been added to understanding of reversible entropy $S^{\text{rev}}$. 
Algorithmic probability $2^{-D^{-}}$ is known and is objectively most 
effective over all computable distributions for a given sequence $x\in 
\mathbb{N}$. Shannon's information has a version in Kolmogorov complexity 
theory but is redundant granted $\mathbb{QKM}$\ \cite{LV}. Other than in the 
trivial data list of \S 3 with measured $T,V,N$, the word ``information'' is 
inappropriate. To use quantity of pattern as a new physical variable is a 
paradigm shift in that there is no ignorance, nor a feasibility 
problem to overcome. The detailed one-particle theories remain true, where 
they apply. Pure thermodynamics remains true. It 
has proved possible to work with intrinsic data and to exclude human
presence. A quantised state in a Langmuir-type site is then objective, \emph{%
where it applies}.%

\section{The wider picture} 

In general, one-particle Schr\"odinger mechanics in large closed volumes 
(including the case of noninteracting particles) must be carefully used. 
For example, by algorithmic correlation quantum chaos in macroscopic 
$\partial V$\ does not admit ML-random energy levels for single-particle 
wave equations \cite{Gutz}. We suggest that in general the deterministic 
single-particle Schr\"odinger problem on macroscopic scales describes 
nonrandom structure.   

There is no helium wave function over $V$ and there is no one such within 
a site. Algorithmic probability for $\mathcal{N}$ atoms over $\mathcal{M}$ 
sites is given by the Langmuir adsorption isotherm as 
$2^{-\mathcal{N}\kappa \mathcal{(M}/\mathcal{N)}}$ and for 
an atom in any particular site in the wall as 
$2^{-\kappa \mathcal{(M}/\mathcal{N)}}$. The two complexities of sites 
are fixed with potentials $-u$ or microscopic lattice openings or the surface 
layer of copper atoms as given. The complexity $\mathcal{N}log\mathcal{N}$ 
is the feed of intrinsic orientations of trajectories in $V$ from given sites 
in a self-consistent balance with incoming trajectories. The decorrelating 
property is simple trapping and scattering on ML-randomly placed sites as in 
\S 3. A Langmuir site is a microscopic structure (see \S 3) with spatial 
scales $\lambda_{th}$ related to temperature by \eqref{11}, directly consistent 
with the action $\lambda_{th}p_{th}=h$. The wall is thermalising with a list of 
detailed exit energies consistent with $\epsilon_{th}={p}^2/2m$. Assuming time
for detailed energy exchange is proportional to a transit time $m{\lambda}/{p}$ 
of a site so that it has sampled copper, recalling that ML-randomness in $V$ demands 
ML-random residence times in sites (\S 3), from a list ${\mathcal{L}_{tau}}$ 
we find detailed actions consistent with ${\epsilon}{\tau}=h/2$. By the two 
actions we find that an atom satisfies marginal conditions for Heisenberg 
uncertainty (HUR) in a site. We do not observe and marginal HUR is an \emph{intrinsic} 
property. Recall the one-by-one exit of atoms from sites. We find a direct 
relationship between Martin-L\"of randomness of a one-by-one accumulated 
list of atom properties, an absolute property of intrinsic data, and a list of 
intrinsic microscopic sites of Heisenberg uncertainty, in the absence of a 
Schr\"odinger wave function. We recall one-by-one ejection 
of particles from the screen of the double-slit 
experiment. Tests for ML-randomness by randomness deficiency would suggest 
detailed mechanics in the screen. Similarly by characterisation of an 
action at marginal Heisenberg uncertainty across the two slits. 
A Schr\"odinger problem defined from source to slits to screen is macroscopic 
and subject to algorithmic correlation, in contract a state at marginal 
uncertainty at the slits would account for ML-randomness at the screen. 

The physical consequences of ML-randomness are in a sense the simple part of
(very) many-body problems - make the assumption then test. If in general
applications an equilibrium state is possible, nonequilibrium dynamics with
relaxation arises with some clues to dynamics from macroscopic observations.
Formally, disorder drops significantly (recall the complexity gap) and
theory of Kolmogorov complexity suggests that simple dynamical models (ie a
few dynamical equations) independent thereof apply. Boltzmann's gas kinetic
theory $\mathbb{BKT}$ applied in $V$\ is a powerful example \cite{B2}. Its
collision integral governs relaxation, is defined in terms of statistical
distribution functions $f(\mathbf{r},\mathbf{v},t)$ in classical phase
space. $\mathbb{BKT}$ in terms of randomness tests on subspaces 
$K(\mathcal{L}_{it}^{\ast })$ rather than $f(\mathbf{r},\mathbf{v},t)$\ are
of interest (perhaps scattering out of a mesoscopic region in equilibrium 
to a turbulent region defined by abrupt drop of complexities?), notably to 
account for the three independent lists. $N$-particle simulations in classical 
mechanics with detailed interaction (eg at the wall) apply before statistics and
consideration of contour maps of ratios of zeros and ones (with other
recursive tests of randomness) in appropriate lists of data at a phase point
would give preliminary insight into Kolmogorov complexity, for example in
turbulent relaxation of a filling gas. The reduction to Navier-Stokes
equations would be of subsequent interest. In the absence of a relaxation
mechanism, thermodynamic concepts in cosmology are risky (formally).

A theory of consistent hierarchies on macro and micro scales is introduced.
It was mentioned in \S 3 that $K(x)\in \lbrack K(y)]$, $x,y\in \mathbb{N}$,
is an equivalence class under Kolmogorov complexity independent of scales of
system size, applying equally to intrinsic data lists in gas theory and for 
example in cosmology. This introduces the possibility of physical analogies 
per scale size. Here we have a consistent hierarchy from micro- to macroscopic 
physics mediated by ML-randomness and Kolmogorov complexity, hierarchy is 
ubiquitous in the observed world and analogy holds, while interconnectedness 
of hierarchies themselves arises. The discovery of $FD_{MN},BE_{MN}$ is a
surprising contribution of $\mathbb{QSM}$, by the perfect property of their
logarithms. It is remarkable that the hidden physics carries an explicit
well-mixed requirement before sampling from spaces over each list, unlike
distributions of academic mathematical statistics where mixing can be
assumed. This physical requirement is of general importance. Observed
regularities are in general to be investigated, the collapse of complexity
\eqref{49} naturally suggesting that deterministic, macroscopic physics
applies.

\end{document}